\documentclass[12pt]{iopart}

\usepackage{array}
\usepackage{graphicx}
\usepackage{dcolumn}
\usepackage{bm}
\usepackage{siunitx}
\usepackage{subfig}
\usepackage{hyperref}
\usepackage{listings}
\usepackage[dvipsnames]{xcolor}
\usepackage{float}
\usepackage{wrapfig}
\usepackage{tcolorbox}
\usepackage{dsfont}
\usepackage{colortbl}
\usepackage{makecell}
\usepackage{longtable}
\usepackage{fontawesome}
\usepackage{censor}
\newcommand\edit[1]{\textcolor{black}{#1}}
\renewcommand{\arraystretch}{1.2}

\definecolor{slate}{HTML}{7493A3}
\definecolor{cloud}{HTML}{A3AFBB}
\definecolor{navy}{HTML}{395464}
\definecolor{sea}{HTML}{668A82}
\definecolor{tan}{HTML}{D6C194}
\definecolor{copper}{HTML}{BF7431}
\definecolor{lightgrey}{HTML}{F5F9F9}
\definecolor{lightgray}{HTML}{949494}
\definecolor{comment}{HTML}{4D31D8}
\definecolor{pyimport}{HTML}{C448CC}
\definecolor{pyfile}{HTML}{B4422C}
\definecolor{pyoptions}{HTML}{447EB4}
\definecolor{pyfunction}{HTML}{60B8C5}
\definecolor{salmon}{HTML}{FDAE7B}
\definecolor{NewBlue}{HTML}{3575D5}

\usepackage{float}

\hypersetup{
colorlinks=true, 
linkcolor=cyan,
urlcolor=cyan,
citecolor=cyan
}

\usepackage{iopams}
\expandafter\let\csname equation*\endcsname\relax
\expandafter\let\csname endequation*\endcsname\relax
\usepackage{mathtools}

\newcommand\mix{\stackrel{\mathclap{\normalfont\mbox{mix}}}{_{j\leq i}}}
\newcommand{\spc}{\phantom{~}}
\newcommand{\sss}{\scriptscriptstyle}
\newcommand{\overhang}{\dimexpr\tabcolsep+0.1pt\relax}
\newcommand{\lruj}{\texttt{\textcolor{copper}{\upshape lrUJ}}}
\newcommand{\ujdet}{\texttt{\textcolor{tan}{\upshape UJdet}}}
\newcommand{\Abinit}{\textsc{Abinit}}
\newcommand{\atompaw}{\textit{atompaw}}
\newcommand{\abinput}[1]{\texttt{\textcolor{slate}{\upshape #1}}}
\newcommand{\abicode}[1]{\texttt{\textcolor{copper}{\upshape #1}}}
\newcolumntype{P}[1]{>{\centering\arraybackslash}m{#1}}
\newcolumntype{L}[1]{>{\raggedright\let\newline\\\arraybackslash\hspace{0pt}}m{#1}}
\newcolumntype{C}[1]{>{\centering\let\newline\\\arraybackslash\hspace{0pt}}m{#1}}
\newcolumntype{R}[1]{>{\raggedleft\let\newline\\\arraybackslash\hspace{0pt}}m{#1}}

\begin{document}

\StopCensoring

\title{Facilities and practices for linear response Hubbard parameters U and J in \textsc{Abinit} }

\author{\censor{L. MacEnulty},$^{1,*}$ \censor{M. Giantomassi},$^2$  \censor{B. Amadon},$^{3,4}$ \censor{G.-M. Rignanese}$^2$ and \censor{D. D. O'Regan,$^1$}}
\address{
 $^1$\blackout{School of Physics, Trinity College Dublin, The University of Dublin, Dublin 2, Ireland}\\
 $^2$\censor{Université} \blackout{Catholique de Louvain, Institute of Condensed Matter and Nanosciences, Chemin des} \censor{Étoiles 8}, \blackout{B-1348 Louvain-la-Neuve, Belgium}\\
 $^3$\censor{CEA, DAM, DIF, }\blackout{91297 Arpajon Cedex, France}\\
  $^4$\censor{Université Paris-Saclay, CEA,} \censor{Laboratoire Matière en Conditions Extrêmes, 91680} \censor{Bruyères-le-Châtel, France}\\
 $^*$Corresponding author
}
\ead{\censor{lmacenul@tcd.ie}}

\vspace{10pt}
\begin{indent}
    \today
\end{indent}

\begin{abstract}
Members of the DFT+U family of functionals are increasingly prevalent methods of addressing errors intrinsic to (semi-) local exchange-correlation functionals at minimum computational cost, but require their parameters U and J to be calculated in situ for a given system of interest, simulation scheme, and runtime parameters. The SCF linear response approach offers ab initio acquisition of the U and has recently been extended to compute the J analogously, which measures localized errors related to exchange-like effects. We introduce a renovated post-processor, the \lruj\ utility, together with this detailed best-practices guide, to enable users of the popular, open-source \Abinit\ first-principles simulation suite to engage easily with in situ Hubbard parameters and streamline their incorporation into material simulations of interest. Features of this utility, which may also interest users and developers of other DFT codes, include $n$-degree polynomial regression, error analysis, Python plotting facilities, didactic documentation, and avenues for further developments. In this technical introduction and guide, we place particular emphasis on the intricacies and potential pitfalls introduced by the projector augmented wave (PAW) method, SCF mixing schemes, and non-linear response, several of which are translatable to DFT+U(+J) implementations in other packages.
\end{abstract}
\vspace{2pc}


\section{\label{Introduction}Introduction}

\Abinit\ \cite{Gonze2020, Romero2020,gonze_abinit_2009,Gonze2008,Gonze2002} is an open-source electronic structure suite developed in the mid-1990s by Xavier Gonze and colleagues. The suite is equipped with a variety of ab initio techniques and the softwares that support them, including, but not limited to, time-dependent density functional theory, dynamical mean field theory, density functional perturbation theory, many-body perturbation theory, electron-phonon calculations, and PAW dataset generation.

Of particular interest here is \Abinit's primary implementation of approximative density functional theory (DFT) and the family of Hubbard-like corrective extensions thereof (referred to here as DFT+U+J). The ground-state DFT scheme relies on a plane-wave self-consistent field (SCF) algorithm, into which developers comprehensively integrated the projector-augmented wave (PAW) method \cite{blochl_projector_1994} in 2008 \cite{torrent_implementation_2008,Amadon2008}.  Shortly thereafter, DFT + U was built on top of this PAW implementation, and programs were developed to calculate its namesake parameter, the Hubbard U, and other corrective objects in situ via linear response \cite{Pickett1998,cococcioni_linear_2005}.

How all these formalisms—DFT+U, PAW, linear response, SCF algorithms—overlap in \Abinit\ is not trivial, and there exists a need for comprehensive, more didactic documentation making explicit the link between the programs and the theory that inspired them \cite{footnote2}. Furthermore, state-of-the-art Hubbard corrective techniques and practices have advanced in the 15 years since the initial implementation of the DFT+U and linear response utilities. For example, the Hund's exchange coupling J and the DFT + U + J functionals have seen more scientific attention than before \cite{himmetoglu2011,ryee2018,orhan2020,Albavera-Mata2022,lambert2023,BLOR2023}, and ground has been made in calculating the Hund J via linear response \cite{Pickett1998,cococcioni_linear_2005,himmetoglu2011,Linscott2018,glenn_thesis}. \Abinit's DFT+U+J linear response implementation and support utilities were due for reevaluation and expansion.

\edit{In addition to the aforementioned aims, this manuscript seeks to illuminate the under-the-hood workings of Hubbard corrective protocol within the PAW formalism, which works in distinct ways among popular electronic structure codes like QuantumESPRESSO \cite{Giannozzi_2009,Giannozzi_2017} and VASP \cite{kresse_ultrasoft_1999}.} 

We address these matters in the current work. In Section \ref{Background}, we provide an overview of the many formalisms involved in \Abinit's DFT+U+J software, including the PAW method (Section \ref{PAW Formalism}), Hubbard corrective protocol (Section \ref{DFT+U_Stuff}), and the linear response method for determining the Hubbard U and Hund's J parameters (collectively referred to herein as the ``Hubbard parameters," the in situ determination of which is described in Section \ref{Linear Response}). The review of this formalism is designed to reinforce our description of the technical implementation of DFT+U+J in \Abinit, which constitutes Section \ref{DFT+U_Abinit} and \edit{includes details on how to invoke various Hubbard parameter-related features via the \Abinit\ input file (Section \ref{Abinit_DFTU}). A closer look at \Abinit's relevant SCF mixing schemes can be found in appendix \ref{Mixing_Schemes}.}

Through the evaluation of \Abinit's linear response faculties in Section \ref{ujdetermination}, we found \Abinit-specific quirks requiring additional clarification as well as evidence of a defect that prompted its renovation and expansion to calculate the Hund's exchange J via first-principles. We describe these renovations in Section \ref{Sec:Explanation} by comparing and contrasting the improved \ujdet\ functionalities with their successor, the \lruj\ post-processor, a user manual of which we document in \ref{Sec:RunningLR}. We then dissect the linear response procedure, following a SCF calculation and documenting the algorithm that transforms potential perturbations to occupancy responses to Hubbard parameters, a chronicle recounted in Section \ref{InternalWorkings}.

A table of contents is provided for ease of navigation.

\newpage
\section*{Table of Contents}
\tableofcontents
\newpage

\section{\label{Background}Background}

\edit{The following sections comprise a description of the formalism on which the \Abinit\ DFT+U(+J) functionalities were built. We begin, in Section \ref{PAW Formalism} with a relatively detailed description of the PAW formalism, introducing its canonical basis functions and  their  origins. This description constructs the necessary foundation on which PAW basis set generators like \atompaw\ are built, providing useful information on how to select a PAW dataset and pseudopotential for use in one's system. Moreover, the PAW and DFT+U(+J) formalisms are particularly entangled when it comes to the use of projectors to isolate the Hubbard subspaces for treatment. The following reiteration of the PAW formalism and its manifestations in DFT is intended to be convenient and accessible reference information for Section \ref{InternalWorkings}, and within that, specifically Section \ref{dmatpuopt}, as we describe how the DFT+U(+J) method is embedded within \Abinit's PAW implementation.}

\subsection{\label{PAW Formalism}PAW Formalism}

In 2008, \Abinit\ developers comprehensively integrated the PAW method, introduced by Peter Blöchl in 1994, into \Abinit\ DFT in order to simplify the numerical treatment of core electron wavefunctions, which are typically encumbered by tight oscillations near the nucleus of atoms. While we summarize the relevant details of the PAW formalism here, we refer the reader to Blöchl's original paper (reference \cite{blochl_projector_1994}) for more information. Details on the PAW implementation in \Abinit\ can be found in references \cite{torrent_implementation_2008,Amadon2008}.

The PAW formalism starts by positing that there exists a true, full all-electron (AE) wavefunction $\Psi$—a Slater determinant of one-electron Kohn-Sham orbitals tightly oscillating about each atomic nucleus—that can be linked to a well-behaved pseudized wavefunction $\tilde{\Psi}$ with few or no oscillations by a linear transformation $\tau$,

\begin{equation}
    \Psi=\tau\tilde{\Psi}.
\end{equation}

\begin{figure}[t!]
    \centering
    \includegraphics[trim={0.5cm 0.5cm 0.25cm 1cm},clip,width=1.0\linewidth]{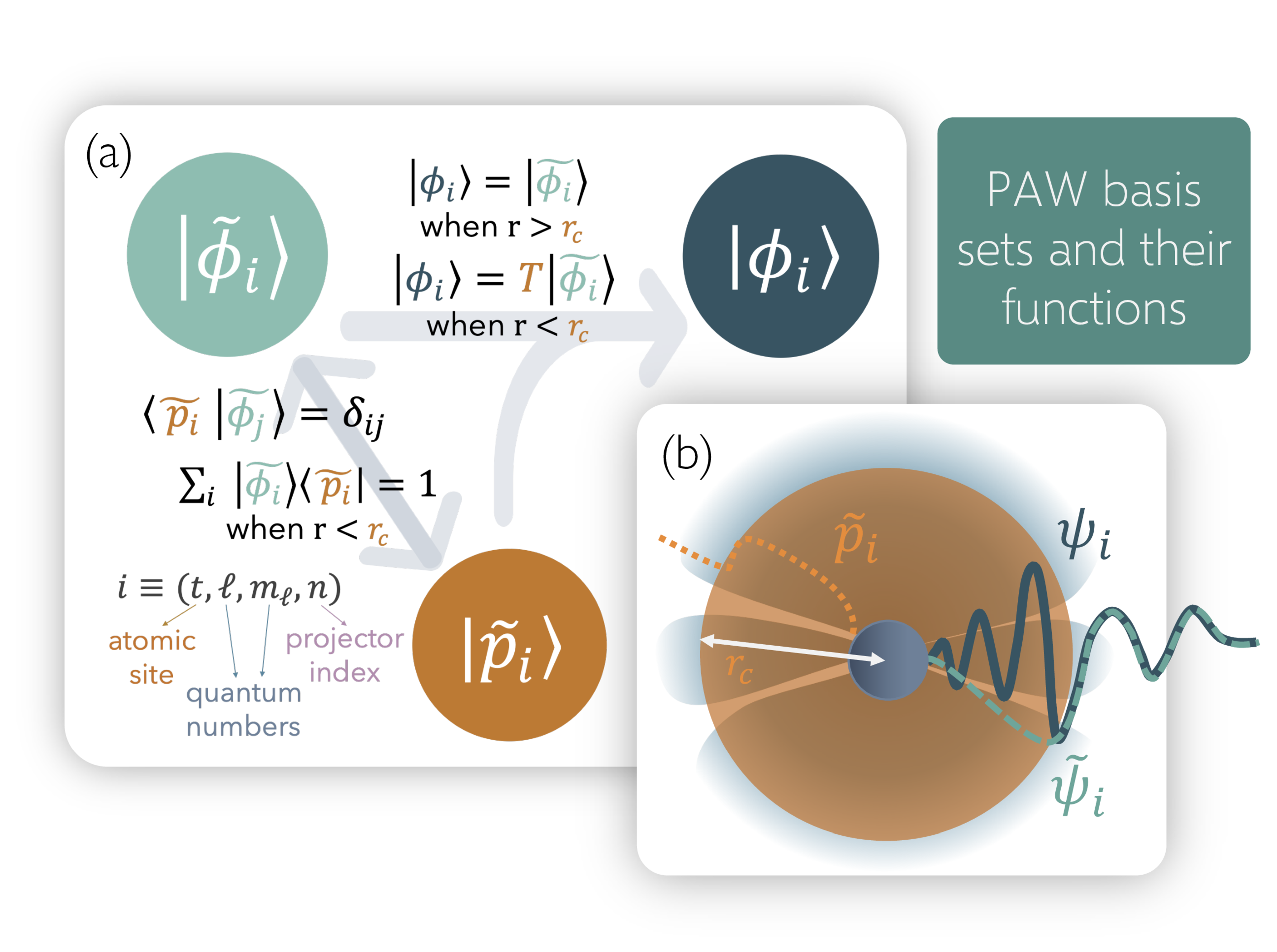}
    \caption{(a) Schematic outlining the mathematical relationship between the three canonical bodies used to construct PAW datasets. (b) The atom-centered augmentation sphere, defined by projector $p_i$, where $i$ refers to the set of four indices $\{t,\ell,m_{\ell},n\}$ inside cut off radius $r_\mathrm{c}$. When $r>r_\mathrm{c}$, the AE wavefunction $\psi_i$ and the pseudo wavefunction $\tilde{\psi}_i$ are equivalent.}
    \label{PAW-bodies}
\end{figure}

With this postulate in hand, we can derive physical quantities using the expectation value of an operator $A$ by sandwiching it between the transformation operator $\tau$ and its Hermitian conjugate $\tau^\dag$ to yield the operator's pseudized counterpart, $\tilde{A}$. Accordingly, the variational principle for the total energy is

\begin{equation}
    \frac{\partial E\left[\tau|\widetilde{\Psi}\rangle\right]}{\partial\langle\widetilde{\Psi}|}=\varepsilon \tau^\dag \tau|\widetilde{\Psi}\rangle
\end{equation}

\noindent such that one may derive the pseudized equivalent of the Kohn-Sham equations. Thus, instead of looking for the ground state of our system in real space, one may seek the ground state energy in this particular pseudospace.

A better-defined transformation operator $\tau$ is necessary to pursue this method any further. As a prerequisite, $\tau$ must modify the smooth, pseudo-valence wavefunction within an atomic region in order to yield the correct nodal structure for the AE wavefunction. This modification is only necessary in the regions closest to the atomic nuclei. We then tailor our treatment specifically to this atom-centered region, spherically symmetric about the nucleus, by defining a cut-off radius, inside which we consider only the smooth, pseudized representation of the AE wavefunction, and outside of which the AE and pseudo wavefunctions are equivalent. Figure \ref{PAW-bodies} illustrates this concept.

In this way, we effectively define an augmentation sphere: a spherically symmetric region of radius $r_\mathrm{c}$ centered around each atomic site in our system. Consider constructing a pseudized wavefunction for a particular orbital with quantum numbers $\ell$ and $m_{\ell}$ on atom $t$. When $r<r_\mathrm{c}$, we must ensure that $\tilde{\psi}_i$ is a well-behaved projection of the all-electron wavefunction $\psi_i$. Moreover, when $r>r_\mathrm{c}$, $\tilde{\psi}_i=\psi_i$. To satisfy these requirements, the transformation operator $\tau$ may be defined as an identity operator plus the sum of atomic orbital-based modifications,

\begin{equation}\label{transform_operator}
    \tau=\mathds{1}+\sum_{i}\left(|\phi_i\rangle-|\widetilde{\phi_i}\rangle\right)\langle\widetilde{p_i}| ,
\end{equation}

\noindent where $\phi_i$ ($\widetilde{\phi_i}$) are the AE (pseudo) partial wave basis functions with which we define the AE (pseudo) wavefunctions $\psi_i$ ($\tilde{\psi}_i$). Here, $i$ refers to the set of four indices $\{t,\ell,m_{\ell},n\}$, representing the atomic site, angular quantum number, magnetic quantum number, and projector index, respectively. The AE partial waves $\phi_i$ may be defined in any way, although an organic choice would be the bound and scattering state solutions to the Schrödinger equation for an isolated atom. In a manner analogous to the wavefunctions they construct, the pseudo partial waves $\widetilde{\phi_i}$ corresponding to each $\tilde{\psi}_i$ are well-behaved projections of $\phi_i$ when $r<r_c$ and identical to $\phi_i$ when $r>r_c$. The objects that allow us to restrict these projections to the pertinent regions of space are the projector functions $\tilde{p}_i$. These are the three mathematical objects intrinsic to the PAW formalism (and those of any ultrasoft pseudo potential method, for that matter). Figure \ref{PAW-bodies} makes explicit the nature of the mathematical duality between these functions. Crucially, note the relation between the pseudo partial waves and the projectors; they are dual to one another, and their projections are normalized inside the augmentation region.

\begin{figure*}[t!]
    \centering
    \includegraphics[width=\linewidth]{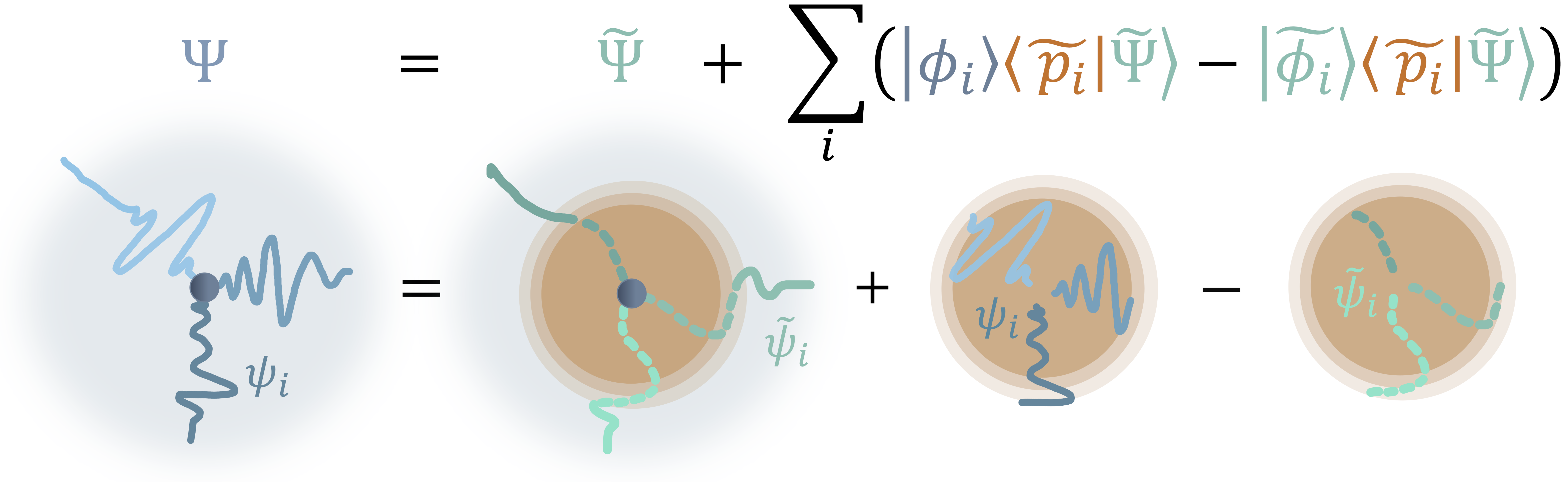}
    \caption{Schematic of the structure of the AE wavefunction in the PAW formalism. In this way, we can restrict our treatment of the cumbersome AE wavefunction to the interior of a particular region.}
    \label{PAW_wfn}
\end{figure*}

We can represent the Kohn-Sham wavefunctions—which can be approximated as plane waves (i.e., waves with wave fronts parallel to flat planes that are azimuthally symmetric about the direction of propagation)—by an infinite sum of spherically symmetric constituents called partial waves (i.e., waves with spherical wave fronts that propagate along a radius emanating from a central point). Such partial waves, also known as spherical waves, are the products of spherical Bessel functions and the spherical harmonics, which are functions of angular momentum $\ell$ and azimuthal quantum number $m_{\ell}$. In the PAW context, therefore, a partial wave refers to a wave that is spherically symmetric about an atom and a function of a given angular momentum $\ell$.

Per equation (\ref{transform_operator}), the atomic orbital based modifications that transform properties into their pseudo-space counterparts comprise the cumulative differences between (i) the projection of the AE partial wave on the augmentation region, and (ii) the pseudo partial waves on the augmentation region. The transformation operator, when applied to the wavefunction, is described visually in figure \ref{PAW_wfn}. Due to the structure of the transformation operator $\tau$, objects such as the density and the energy are formulated analogously.

\subsubsection{\label{PAW_dataset}\textcolor{lightgray}{Construction of PAW datasets}}

The three PAW basis sets $\psi_i$, $\tilde{\psi}_i$, and $\tilde{p}_i$, are typically read into one's chosen DFT program from a file, called a PAW dataset, not so unlike a pseudopotential. A PAW dataset to be read into \Abinit\ should adhere to either XML format (extension \texttt{.xml}) or the Abinit proprietary format (extension \texttt{.abinit}). These datasets will typically include the following information.

\begin{enumerate}
	\item AE and PS partial waves, $\psi_i$ and $\tilde{\psi}_i$, and projector functions, $\tilde{p}_i$, for all valence orbitals
	\item AE core charge density, $n_c$
	\item PS core charge density, $\tilde{n}_{c}$
	\item PS valence charge density, $\tilde{n}^1$
	\item A local ionic pseudopotential, $\tilde{\nu}_\mathrm{loc}$ (see appendix \ref{PAW_pseudopotentials})
	\item Information for the construction of the compensation charge, $\hat{n}$, (i.e., a shape function)
\end{enumerate}

There exist a variety of publicly available programs to generate these datasets, the one with ties to \Abinit\ being \atompaw\ \cite{holzwarth_projector_2001,holzwarth_notes_nodate,JTH2014}. 

These algorithms start by defining the AE partial waves as descriptions of all orbitals of the element under scrutiny. As mentioned earlier, the eigenfunctions of the solution to the Schrödinger equation for the isolated atom are a good starting point as they are partial waves, thus comprising a sum of products—radial functions (typically of polynomial form, although other options are open for use) multiplied by spherical harmonics. DFT calculations are performed using the exchange-correlation potential, $\nu_\mathrm{xc}$, of choice to obtain the AE basis functions defined on a radial grid of variable mesh density. For this reason, PAW datasets are categorized in terms of the associated XC functional (e.g., PAW-LDA, PAW-PBE). The \atompaw\ algorithm, specifically, mandates that the partial waves be defined as the atomic eigenfunctions resulting from these calculations.

We assume here that the electronic states resulting from this calculation can be separated into core orbitals—to be ``frozen" only in the sense that they are grouped with the nucleus and not represented individually by PAW basis sets—and valence orbitals, which will be represented individually in the PAW basis \cite{JTH2014}. In principle, the frozen core approximation that inspired this treatment would necessitate the core electron density remain unchanged from that of the relaxed isolated atom despite immersion in different environmental potentials and configurations. By drawing more or less orbitals out of the core and into the valence, the accuracy of this frozen core approximation may be tailored to the system at hand. The number of PAW basis functions, used to describe these valence orbitals, follows from this decision.

In practice, and in \Abinit, a soft-core scheme is adopted to restore the core electrons' involvement in and response to the physics of the system. A pseudo density representing the nucleus and core electrons, $\tilde{n}_\mathrm{Zc}$, is used to generate its namesake share of the Hartree potential, and a soft-core density, $\tilde{n}_c$, contributes to calculations of any non-linear core corrections \cite{hine2017}. Additionally, a compensation charge density, $\hat{n}$, is introduced to restore the correct multipole moments of the AE charge density $n^1+n_\mathrm{Zc}$, evaluated on a radial grid and exclusively inside the augmentation region \cite{kresse_ultrasoft_1999}.

Often, only one or two partial waves are necessary to accurately describe a valence orbital's angular momentum, as the partial wave expansion of orbitals rapidly converges. At least one partial wave constructing a PS plane wave will represent a bound electronic state of the atom. Often, when there are two partial waves representing a subspace, the PAW dataset will feature one bound and one unbound electronic state.

The pseudo partial waves and the associated projectors are then constructed from one of many pseudization scheme designed to ensure that these bodies fulfill the requisites established in earlier in this section. The PS partial waves are solutions to the PAW Hamiltonian, which features a screened, pseudized local potential that is equal to the AE atomic potential outside of a radius $r_\mathrm{pp}$. One option for the pseudization scheme is that of Blöchl, in which the pseudized basis functions are solutions to the non-relativistic Schrödinger equation of an isolated atom immersed in a pseudopotential defined for each AE partial wave.  In other words, the projector functions are chosen first, and the PS basis functions are derived \cite{holzwarth_notes_nodate}. This procedure is described more in Section VI B of reference \cite{blochl_projector_1994}. Yet another option involves the RRKJ optimization scheme \cite{RRKJ1990}, which represents the pseudo wavefunction as a sum of two Bessel functions.

The pseudization scheme with which we are most concerned, for reasons that will be made known in Section \ref{dmatpuopt} is the so-called Vanderbilt \cite{vanderbilt_soft_1990} scheme implemented in \atompaw. Via this scheme, the pseudo wavefunctions are eighth-degree polynomials inside the augmentation region,

\begin{equation}
    \tilde{\psi}_i(r)=r^{\ell+1} \sum_{k=0}^4 c_k r^{2k},
\end{equation}

\noindent from which the basis and projector functions are deduced by fitting the five coefficients $c_k$ such that $\tilde{\phi}_i=\phi_i$ in the vicinity of $r_c$. The Vanderbilt scheme thus ensures that (a) $\tilde{\psi}_i(r)$ is an eigenfunction of the atomic PAW Hamiltonian; and (b) the $\tilde{\psi}_i(r)$ satisfies the generalized norm-conserving condition $Q_{ij}=0$, where 

\begin{equation}\label{Qij}
    Q_{ij}=\int_{r<r_{\mathrm{c},i}}{[\phi_i^\ast\left(\mathbf{r}\right)\phi_j\left(\mathbf{r}\right)-\widetilde{\phi}_i^\ast\left(\mathbf{r}\right) \widetilde{\phi}_j\left(\mathbf{r}\right)]\ d\mathbf{r}}.
\end{equation}

\noindent and $j$ represents a set of quantum numbers $\{t,\ell,m_{\ell},n\}$ distinct from that of $i$. The integrand of $Q_{ij}$ is known as the pseudized augmentation function between two partial waves $i$ and $j$. For more details on this scheme, we refer the reader to references \cite{holzwarth_notes_nodate,vanderbilt_soft_1990}.

Following the pseudization process, the pseudo basis functions and their accompanying projectors are orthogonalized according to a chosen orthogonalization scheme, such as the Gram-Schmidt scheme or Vanderbilt's own scheme, also published in reference \cite{vanderbilt_soft_1990}.

With the AE basis functions, pseudo basis functions and projector functions defined, PAW dataset generators output their values on a radial grid in the form of a pseudopotential in an increasing number of formats compatible with popular electronic structure theory codes. Predefined, open-source, and tested PAW datasets for many elements can be acquired from, for example, the \href{http://users.wfu.edu/natalie/papers/pwpaw/man.html}{\atompaw}\ or \href{http://www.pseudo-dojo.org/}{PseudoDojo} \cite{JTH2014,pseudodojo2018} websites, among others.

\subsection{\label{DFT+U_Stuff}DFT+U+J}

Inspired by the Hubbard model \cite{Hubbard1963}, DFT+U+J \cite{anisimov_band_1991,anisimov_density_1991,dudarev1998,anisimov_density_1993,himmetoglu2014} offers treatment of self-interaction and static-correlation errors in highly localized electronic subspaces while minimizing additional computational expense \cite{oregan_linear_2012}. There are a variety of functionals occupying the Hubbard-like corrective class, ranging from the recognizable and widely implemented Dudarev (DFT+U$_\textrm{eff}$=U-J) \cite{dudarev1998}, Himmetoglu (DFT+U+J) \cite{himmetoglu2014}, and Liechtenstein \cite{Lichtenstein1995} functionals, to the ultra-modern BLOR functional \cite{BLOR2023}, derived to explicitly address the flat-plane condition. The treatment provided by the additional Hubbard U and Hund's J terms is prescribed exclusively for subspaces that require numerical attention in excess of those for which the base XC functional is descriptively sufficient.

We assume that in consulting this technical report at all, the reader is already operationally familiar with DFT+U, if not DFT+U+J, as a method. Furthermore, the modifications we present here did not extend to the implementation of the Hubbard functionals themselves, only the calculation of their parameters. For this reason, rather than provide an overall review, we train our focus on the linear response determination of the Hubbard U and Hund's J in addition to the specifics of how DFT+U+J is implemented in \Abinit. (See Section \ref{DFT+U_Abinit}). For more information on the formalism and relative advantages of DFT+U+J method, we redirect the reader to references. \cite{Amadon2008,BLOR2023,Linscott2018,anisimov_band_1991,anisimov_density_1991,dudarev1998,anisimov_density_1993,himmetoglu2014,Lichtenstein1995,Anisimov_1997}.

\subsection{\label{Linear Response}Linear Response determination of the Hubbard parameters}

In the relevant literature, the Hubbard U parameter for a valence subspace is most often determined semi-empirically. That is, the parameter space is swept and a U value chosen for its ability to get a particular DFT-derived property closer to some more concrete benchmark. Otherwise, the U and J parameters are reappropriated from similar studies, in perpetual and serial reuse. However, since these parameters are ground state properties of subspaces described by a particular XC functional (and moreover specific to code, pseudopotential, and other runtime convergence parameters), they are inherently non-transferable on the one hand, but also derivable from first-principles on the other. 

Following Pickett \textit{et al.}'s \cite{Pickett1998} lead, Cococcioni and de Gironcoli picked up and developed a linear response-based protocol for calculating the Hubbard U in situ \cite{cococcioni_linear_2005}. \Abinit\ researcher Donat Adams, alongside Bernard Amadon and Silke Biermann, developed an \Abinit\ utility, manifested in the PAW formalism, that determines the strength of the Coulombic repulsion, the Hubbard U, and other metrics via linear response. Details of this utility can be found in appendix \ref{AppendixSec:ujdet}. The homologous linear response protocol for the Hund's J was published by Linscott \textit{et al.} in 2018 \cite{Linscott2018}, taking inspiration from the earlier exploration of the same by Himmetoglu \textit{et al} \cite{himmetoglu2011}. We refer the reader to these articles for the mathematical formalism and theory.

Formulated practicably, the SCF linear response procedure for calculating a scalar Hubbard U parameter is as follows.

\begin{enumerate}
    \item Serially apply several (preferably more than three) small perturbations $\pm\alpha$ in equal magnitude to both the up and down spin channels of the external potential of the chosen error-afflicted subspace.
    \item Once the potential perturbation $\alpha$ is applied to the treated subspace, the charge occupations on the spin-up $n^\uparrow$ and spin-down $n^\downarrow$ channels of that atom and the surrounding atoms change in response. The change in occupation as a direct result of this potential perturbation is nothing more than the non-interacting response function, $\chi_0$, which is harvested here, after the first SCF iteration but before the density and Hamiltonian are updated to begin a new iteration. Therefore, extract the up and down subspace occupations of the subspace, $n^\uparrow_0$ and $n^\downarrow_0$, after the first self-consistency iteration.
    \item The perturbation is screened, its charge reorganized to compensate for the disturbance once again until, after the last self-consistent iteration, it reaches an equilibrium state. At the end of the SCF cycle, then, extract the up and down subspace occupations, $n^\uparrow$ and $n^\downarrow$. The derivative of this equilibrium occupation with respect to the perturbation magnitude is the interacting response function, $\chi$.
    \item Perform a linear (or higher-order polynomial) regression to the collected data sets and differentiate at $\alpha=\beta=0.0$ eV to find the slopes of the response functions, $\chi_0=\frac{d(n^\uparrow_0+n^\downarrow_0)}{d\alpha}$ and $\chi=\frac{d(n^\uparrow+n^\downarrow)}{d\alpha}$.
    \item Insert the response functions into the following equation to acquire U.
        \begin{align}\label{Utechnical}
            U&=\frac{d\alpha}{d(n_0^\uparrow+n_0^\downarrow)}-\frac{d\alpha}{d\left(n^\uparrow+n^\downarrow\right)} \nonumber \\
            &=\chi_0^{-1}-\chi^{-1}.
        \end{align}
\end{enumerate}

The extension to polynomial regressions in step (iv) accounts for the fact that the response behavior is not always linear. Figure \ref{Plot:LinearResponsePlot} partially demonstrates this for the Ni $3d$ orbitals in a ferromagnetically ordered NiO system. Based on the visuals alone, one can see the data demonstrate a noticeable degree of curvature. The use of a linear regression on this response is not entirely justified. A system demonstrating exceptionally ill-behaved linear response, wherein a third-order polynomial or higher would be needed to accurately fit the data, can be seen in figure 2 of reference \cite{MacEnulty2023}. Linearity is expected in the limit of small perturbations, but this region is not always accessible if one wants to amplify the signal-to-noise ratio. If the perturbations are too large, one can expect some non-linear behavior, or even asymmetry across the zero-perturbation axis, particularly if the system has a shallow energy landscape.

Ideally, U is calculated for a cell of infinite size such that the perturbed subspace is isolated from its periodic images. Since this is unfeasible computationally, U must be converged with respect to an increasing number of atoms, ideally organized into a roughly cubic supercell to isotropically distribute the effect of the perturbation.

The Hund's coupling J parameter is calculated analogously \cite{Linscott2018,anisimov_band_1991}. Instead of monitoring the change in total subspace occupancy as a function of the applied perturbation $\alpha$, however, Hund's J monitors changes in magnetization M, or difference between the up and down spin occupancies (i.e., M $=n^\uparrow-n^\downarrow$), in response to perturbations $\pm\beta$ applied in positive magnitude to the spin-up potential and in negative magnitude to the spin-down potential. Furthermore, the sign convention in the calculation of J is the opposite of that for the Hubbard U (i.e., a positive J corresponds to the curvature of the total energy with respect to fractional magnetization, minus its non-interacting analogue, demonstrating concavity). Formally,

\begin{align}\label{Eq:Jtechnical}
\textrm{J}&=\frac{d\beta}{d\left(n^\uparrow-n^\downarrow\right)}-\frac{d\beta}{d(n_0^\uparrow-n_0^\downarrow)} \nonumber \\
&=\chi_\textrm{M}^{-1}-\chi_{0_\textrm{M}}^{-1}.
\end{align}

\vspace{1cm}

\begin{center}
\begin{tcolorbox}[width=0.8\linewidth,colback={lightgrey},title={NOTE: Equivalency of Unscreened Response Functions for $\alpha$ and $\beta$ Perturbations},colbacktitle=sea,coltitle=white]
    It is important to note, for verification purposes, that the unscreened response matrices $\chi_0$ and $\chi_{0_\textrm{M}}$ are equivalent. That is, for the same material subspace, we can show that $\chi_0=\chi_{0_\textrm{M}}$. The proof of this can be found in Appendix A of reference \cite{MacEnulty2023}.
\end{tcolorbox}
\end{center}

\section{\label{Abinit_DFTU}Implementation of DFT+U+J in \textsc{\Abinit}}

This and following sections are color- and font- coded for clarity. Mutable variables found in the \Abinit\ input file are displayed in \abinput{blue code} text. Immutable, internal \Abinit\ variables, functions and subroutines are displayed in \abicode{orange code} text.

\subsection{\label{DFT+U_Abinit}Running DFT+U in \Abinit}

The DFT+U formalism is built into \Abinit's PAW functionality. As of version 9, DFT + U and PAW are inseparable in \Abinit\, and its users have no choice but to use PAW datasets as pseudopotentials when administering a correction via the Hubbard functionals. Moreover, the Hubbard functional implementation and related utilities in \Abinit\ are, for the moment, restricted to the case of collinear magnetism (i.e., when the variable \abinput{nspinor}\texttt{=1}).

When DFT+U and PAW are simultaneously activated, the total energy becomes
\begin{equation}\label{DFTU_functionalform}
    E_\textrm{DFT+U}=E_\textrm{DFT}+E_{ee}-E_\textrm{dc}
\end{equation}
\noindent where $E_\textrm{DFT}$ is the standard DFT energy functional, $E_{ee}$ is the electron-electron interaction energy expanded in equation (1) of reference \cite{Amadon2008}, and $E_\textrm{dc}$ is the double-counting term, which corrects for the interaction already encompassed in $E_\textrm{DFT}$.

DFT+U is activated via the \abinput{usepawu} input variable, a single integer which may adopt several non-zero values, each corresponding to the treatment of the double-counting term. If \abinput{usepawu}\texttt{ = 0}, DFT+U is unactivated. If \abinput{usepawu}\texttt{ = 1}, the double-counting term is assessed via the Full Localized Limit formulation proposed by Anisimov \textit{et al.} \cite{anisimov_band_1991}, which takes on the following form,

\begin{equation}\label{FLL}
    E_\textrm{dc}=\frac{U}{2}\ N\left(N-1\right)-\frac{J}{2}\sum_{\sigma}{N^\sigma\left(N^\sigma-1\right)}.
\end{equation}

When only the Hubbard U is defined via input variable \abinput{upawu}, the Hund's J is assumed to be 0.0, and equation (\ref{FLL}) inserted in equation (\ref{DFTU_functionalform}) becomes the Dudarev functional (DFT+U$_\mathrm{eff}$) \cite{dudarev1998}. When the Hund's J is set to a non-zero value via \abinput{jpawu,} equation (\ref{FLL}) contributes to a Hubbard corrective protocol sometimes called the Liechtenstein \cite{Lichtenstein1995} DFT+U+J functional. Its double-counting expression is derived from a reference system that assumes the diagonal elements of the diagonalized occupation matrix are integers. Similarly, this expression is evaluated if \abinput{usepawu}\texttt{ = 4}, except it is done so without spin polarization in the exchange-correlation functional \cite{Chen2016}. If \abinput{usepawu}\texttt{ = 2}, the Around Mean Field double counting expression, found in equation (7) of reference \cite{Czyzyk_1994}, is evaluated. Other options for the \abinput{usepawu} variable exist, which are related to DMFT and GW methods. For the standard DFT protocol with Hubbard corrections, use \abinput{usepawu}\texttt{ = 1}, the Full Localized Limit.

Declaration of the \abinput{usepawu} compels \Abinit\ to read three more input values: \abinput{lpawu}, \abinput{upawu} and \abinput{jpawu}. The variable \abinput{lpawu} accepts an array of integers of length \abinput{ntypat} (the number of types of atoms) to determine on which atomic subspaces we will apply U and J values. If \abinput{lpawu} is negative, no Hubbard parameters are applied. If \abinput{lpawu} is positive, \Abinit\ will apply a Hubbard U and Hund's J to the atomic subspace indexed by the angular quantum number of the subspace (e.g., \abinput{lpawu}\texttt{ = 2} applies it to \textit{d} orbitals, \abinput{lpawu}\texttt{ = 3} to \textit{f} orbitals). The U and J can be applied to any orbitals, including \textit{s} orbitals. The variables \abinput{upawu} and \abinput{jpawu}, subsequently, define respectively the Hubbard U and Hund's J parameters to be applied to those subspaces. By default, \abinput{upawu} and \abinput{jpawu} are read in atomic units but can be specified in other units of energy, notably eV.

Optional variables for \Abinit's DFT+U implementation include \abinput{usedmatpu} and \abinput{dmatpawu}, which work together to allow the user to propose an initial density matrix to facilitate \Abinit\ in finding the DFT+U ground state.

\section{\label{ujdetermination}Determination of the Hubbard parameters in situ in \Abinit}

There are two ways to determine the Hubbard parameters in situ with \Abinit: linear response (\lruj\ or \ujdet), or cRPA. The cRPA protocol is beyond the scope of the present article, but the interested reader may take a look at the \href{https://docs.abinit.org/tutorial/ucalc_crpa/}{cRPA \Abinit\ tutorial} in addition to references \cite{Amadon2014,Aryasetiawan2004} to get started.

Prior to \Abinit\ version 9.9, only the \ujdet\ internal and post-processing utilities existed as a means of calculating the Hubbard parameters via linear response in \Abinit. In 2022, users alerted \Abinit\ to some inconsistencies in its implementation. These inconsistencies are explained in appendix \ref{AppendixSec:ujdet}. For technical reasons, however, these issues could not be easily remedied, and the decision was taken to decommission the \ujdet\ post-processing utility and to renovate its internal functionality. 

As of version 9.10, the \Abinit\ DFT suite is equipped with both the renovated \ujdet\ utility in addition to a new post-processing tool, the Linear Response U(J) (\lruj) utility, which is built upon the same core \ujdet\ programming. Most of \ujdet's data processing functionalities have been preserved throughout this renovation. However, we emphasize that the functionalities of \ujdet\ and \lruj\ serve distinct purposes and implement different levels of theory, which we discuss further in the following sections.

Although older versions of \Abinit\ preserve the \ujdet\ deprecated internal functions and post-processing utility, their use is strongly disadvised for the reasons outlined in appendix \ref{AppendixSec:ujdet}.

\begin{table}[t]
\footnotesize
\setlength\tabcolsep{5pt}
\setlength{\arrayrulewidth}{0.4mm}
\bgroup
\arrayrulecolor{white}
\def\arraystretch{1.5}
{\fontfamily{cmss}\selectfont
\begin{tabular}{P{0.02\linewidth}|P{0.45\linewidth}|P{0.45\linewidth}|}

\rowcolor[HTML]{EDEDED}[\overhang] 
{\cellcolor{white}}&
\centering{\cellcolor{tan}{\normalsize \texttt{\textcolor{white}{UJdet}}}} &
\centering\arraybackslash{\cellcolor{copper}{\normalsize \texttt{\textcolor{white}{lrUJ}}}} \\ \hline

\rowcolor[HTML]{EDEDED}[\overhang]
{\cellcolor{navy}\textcolor{white}{1}} & Embedded in \Abinit\ core routine & Post-processor \\ \hline

\rowcolor[HTML]{EDEDED}[\overhang]
{\cellcolor{navy}\textcolor{white}{2}} & Two-point linear regression & 3+ point polynomial (variable degree) regression \\ \hline

\rowcolor[HTML]{EDEDED}[\overhang]
{\cellcolor{navy}\textcolor{white}{3}} & $\boldsymbol{\chi}$ and $\boldsymbol{\chi_0}$ responses treated as matrices; interatomic response monitored; matrices augmented by total system charge & $\chi$ and $\chi_0$ responses treated as scalars \\ \hline

\rowcolor[HTML]{EDEDED}[\overhang]
{\cellcolor{navy}\textcolor{white}{4}} & Supercell extrapolation scheme & RMS Error analysis \\ \hline

\rowcolor[HTML]{EDEDED}[\overhang]
{\cellcolor{navy}\textcolor{white}{5}} & Atomic Sphere Approximation projector extensions/normalizations & Outputs \texttt{*LRUJ.nc} NetCDF files with details of perturbative run \\ \hline

\end{tabular}}
\egroup
\caption{\label{Tab:lrujvsujdet}Comparison of preserved and renovated \Abinit\ linear response functionalities.}
\end{table}

\subsection{\label{Sec:Explanation}Clarification of available linear response utilities}

The primary differences between the \lruj\ and \ujdet\ as implemented in current versions of \Abinit\ are outlined in table \ref{Tab:lrujvsujdet}, where item (2) highlights the most obvious difference between the two: the number of data points used to compute a linear regression of the response functions $\chi$ and $\chi_0$. The \ujdet\ utility uses only two points: the unperturbed case—in which the perturbation applied is zero and the subspace occupations are those of the ground state—and one perturbed case, in which the potential perturbation is equal in magnitude to the value of input variable \abinput{pawujv}.

By contrast, the \lruj\ utility requires, at minimum, three data points (one unperturbed case and at least two perturbations) to conduct a distinct regression analysis. With $n$ data points, the \lruj\ utility computes not only a linear regression of the response functions $\chi$ and $\chi_0$, but all polynomial regressions up to degree $n - 2$. Furthermore, the \lruj\ utility conducts RMS error analysis on the fits and factors that into an approximative RMS error on the resulting Hubbard parameters.

\begin{center}
\begin{tcolorbox}[width=0.8\linewidth,colback={lightgrey},title={NOTE: Insufficiency of two-point regression},colbacktitle=sea,coltitle=white]
    Adequate sampling of the $N(\alpha)$ (in the case of the Hubbard U) or $M(\beta)$ (for the Hund's J) terrains is crucial. It is arguable that the two-point interpolation scheme implemented in the \Abinit\ \ujdet\ utility, as demonstrated by the difference taken in equations (\ref{chi0_abi}) and (\ref{chi_abi}), is insufficient in terms of sampling. Furthermore, no insight into the regression error may be achieved using a two-point regression. For \edit{these reasons}, the authors recommend performing multiple perturbations.
\end{tcolorbox}
\end{center}

Another crucial difference between the two utilities is Item (3) in table \ref{Tab:lrujvsujdet}: the \ujdet\ utility treats the response functions as matrices, whereas the \lruj\ utility treats them as scalars. This means that the \ujdet\ Hubbard parameters are, to some degree, informed by the Hubbard interactions on and between the other atomic subspaces of the system as well as the total charge bath.

The protocol is expanded upon in reference \cite{cococcioni_thesis}, wherein an extrapolation scheme aiming to accelerate the determination of the Hubbard parameters is proposed. This scheme involves (a) augmenting the response matrices (collecting the response functions while moving the site of perturbation) with the negative of their total response to enforce charge neutrality, and (b) capitalizing on the assumption that the occupancy response to the potential perturbation attenuates for atoms further away from the site of the perturbation \cite{cococcioni_linear_2005}.

\begin{center}
\begin{tcolorbox}[width=0.8\linewidth,colback={lightgrey},title={NOTE: Using a supercell},colbacktitle=sea,coltitle=white]
    All atoms in the unit cell of a bulk material are bound to each other by symmetry. Altering the subspace potential and/or occupations of one atom will alter the subspaces potentials of the other symmetry-related atoms of the same species, both inside the explicitly defined cell and among their mirror images. \\
    \spc\spc This ripple effect is counterproductive to the perturbation cardinal to linear response; the occupancy of the perturbed subspace must be allowed to evolve separately to both its mirror images and other atoms of the same species. \\
    \spc\spc To ensure that \Abinit\ isolates the perturbed atom from its mirror images, there’s really only one option: perform linear response on the biggest supercell possible. It’s not ideal, and protocols have been implemented to approximatively circumvent this requisite (such as the supercell extrapolation scheme). However, the authors here suggest performing linear response on an actual supercell of the material.
\end{tcolorbox}
\end{center}

By contrast, the \lruj\ utility provides the scalar Hubbard parameters, informed only by the change in occupancy on the perturbed subspace. \edit{Certainly, a more scientific investigation, one beyond the scope of this technical note, is needed to determine which protocol (the scalar or matrix treatment) yields faster convergence with respect to supercell size, and, more abstractly, which is better suited for addressing the self-interaction error and static correlation error within isolated subspaces. In the present work, we do not offer any particular recommendation or endorsement either way, that being beyond the scope of the work. For now, the \lruj\ post-processor does not have the infrastructure needed to perform the matrix response protocol, which would require performing polynomial regression analysis on all elements of the response matrices, a much more computationally complex, taxing, and of yet uncharted endeavor. For those wishing to undergo such an endeavor, we provide details on how to access all of its necessary components in appendix \ref{ujdet_calculation}.}

For all other purposes, it can be said that \lruj\ offers a simplified data processing procedure to that of \ujdet, provided that the user commits to more than three LR data points (i.e., at least two separate DFT runs). By design, these data points can be run in parallel, and so the use of the \lruj\ utility over the \ujdet\ utility is strongly encouraged.

\subsection{\label{Sec:RunningLR}Running Linear Response with \emph{\lruj}}

The explanation that follows is a more detailed version of the corresponding \Abinit\ \href{https://docs.abinit.org/tutorial/lruj/}{tutorial}, \edit{conducted now for the Hund's J parameter as opposed to the Hubbard U. The reasoning for including such slightly redundant information is three-fold. First, in the same way that the official Abinit tutorial describes the calculation and provides example output of the Hubbard U parameter, it is fitting to provide example output files for the same system, but for the calculation of the lesser known yet increasingly emphasized Hund’s J parameter. Furthermore, we take advantage of the visualization opportunities presented by this journal publication to comprehensively label all output variables pertaining to the linear response functionalities implemented in Abinit.}

\edit{Second, the tutorial is extended to include, in Section \ref{PlottingLR}, a proof on the data-processing modifications necessary to correctly plot the linear response data coming from Abinit, and \edit{(in appendix \ref{AbiPy})} a step-by-step guide of the AbiPy functionalities available to plot the results from the lrUJ utility. Neither of these is included in the Abinit documentation.}

\edit{And third, Section \ref{InternalWorkings} on the implementation of the Abinit linear response utilities relies heavily on the chronological outline and input variables best introduced in tutorial format. The current format, starting with a reiteration of the tutorial and ending with a description of the hard-coding mechanisms in Abinit, was designed to make explicit, in an organized fashion, how Abinit’s input variables are connected to its output variables.}

The linear response procedure can be carried out in three steps:

\begin{enumerate}
    \item[1.] Run a ground state \Abinit\ calculation of your supercell to generate \texttt{WFK} files.
    \item[2.] Run a series of perturbative \Abinit\ calculations to generate \texttt{LRUJ.nc} files.
    \item[3.] Execute the \lruj\ post-processing utility.
\end{enumerate}

\subsubsection{\label{Step1LRUJ}\textcolor{lightgray}{Ground state calculation and generation of \emph{\texttt{WFK}} files}}

We need to establish a ground state system whose subspace potential we can perturb. For all intents and purposes, this should be your ordinary DFT calculation, aside from a few minor modifications to the input file.

\begin{center}
\begin{tcolorbox}[width=0.8\linewidth,colback={lightgrey},title={NOTE: Which atom should be perturbed?},colbacktitle=sea,coltitle=white]
    The short answer is, any atom, at least for \lruj. If you wish to avail of the \ujdet\ functionalities such as the supercell extrapolation scheme, the atom on which you intend to apply the linear response  perturbation MUST  be  the  first atom listed in \abicode{xred}. See appendix \ref{ujdet_calculation} for more details.
\end{tcolorbox}
\end{center}

First, we specify as a separate species the atom whose subspace we wish to apply a potential perturbation. This will alert \Abinit\ that we want to allow the perturbed subspace to vary its external potential independently to its kin atoms in the cell. To this end, we increase \abicode{ntypat} by 1 and adjust the parameters \abinput{typat}, \abinput{znucl}, \abinput{lpawu}, \abinput{upawu}, \abinput{jpawu}, \abinput{pseudos}, and all other variables dependent on \abinput{ntypat}, to reflect that change. This will remain true for Step (2), as well.

\begin{center}
\begin{tcolorbox}[width=0.8\linewidth,colback={lightgrey},title={NOTE: Another way to break symmetries},colbacktitle=sea,coltitle=white]
    By specifying the perturbed atom as a separate species, \Abinit\ will only harvest the changes in occupation of the perturbed atom. This information is sufficient for the \lruj\ procedure, but not for \ujdet. To avail of the supercell extrapolation technique, you will need to set the symmetry relations, \abinput{symrel}, explicitly. The symmetry relations are specified in terms of $3\times3$ matrices, and they represent in some form the Wyckoff positions of the atoms. Making these explicit in the input will tell \Abinit\ that (a) the perturbed atom should vary independently to its kin, and (b) it should still collect occupation information for all atoms containing the subspace to be treated, not just that of the perturbed atom. The \ujdet\ utility then uses these interatomic response matrix elements to inform its Hubbard parameters. \\
    \spc\spc You can generate these symmetries in a separate run, wherein you specify the atom upon which the perturbation is to be applied as a different species, in the same way described in this \lruj\ tutorial. From the output, you read the number of symmetries (\abinput{nsym}), the symmetry operations (\abinput{symrel}), and the translation vectors (\abinput{tnons}).
\end{tcolorbox}
\end{center}

In what follows, we assume that the input U and J values are zero. To do this, you can either set all values in \abinput{upawu} and \abinput{jpawu} to 0.0, or you can simply deactivate DFT+U by setting \abinput{usepawu=0}. Crucially, make sure \abinput{prtwf} is set to 1 so that the \texttt{WFK} file is printed.

Once you have all aspects of your ground state run assembled, launch \Abinit\ with the input file to acquire your \texttt{WFK} file.

\subsubsection{\label{Step2LRUJ}\textcolor{lightgray}{Perturbative calculations and generation of \emph{\texttt{LRUJ.nc}} files}}

Once we have our reference wavefunctions, we can start the linear response procedure. We will take advantage of \Abinit's dataset functionality to iteratively apply perturbations of varying strength to our chosen subspace. For now, we describe the input variables needed to perform one such perturbation.

Building on top of the input file used in Section \ref{Step1LRUJ}, we further activate linear response with one input parameter: \abinput{macro\_uj}. This parameter's integer value, in combination with the value \abinput{nsppol} (the number of independent spin channels), determines how the local potential perturbation is applied and the subsequent changes in occupancy harvested. These options are organized in table \ref{macro_uj_options}. The options \abinput{macro\_uj} \texttt{= 1} and \abinput{nsppol} \texttt{= 1} represent the non-spin-polarized case, where total occupations are double those of one spin channel. Importantly, note that both \ujdet\ and \lruj\ are implemented exclusively for the non-collinear case as the linear response theory governing its implementation requires further consideration \cite{binci2023,moore2022}.

\begin{table}[t]
\centering
\arrayrulecolor{white}
\setlength{\arrayrulewidth}{0.4mm}
{\fontfamily{cmss}\selectfont
\begin{tabular}
{P{0.17\textwidth}|P{0.13\textwidth}|P{0.17\textwidth}|P{0.09\textwidth}|P{0.09\textwidth}|P{0.18\textwidth}}

\rowcolor[HTML]{668A82}[\overhang]
{\cellcolor{white}} & \multicolumn{5}{c}{\textcolor{white}{Possible Combinations}} \\ \hline

\rowcolor[HTML]{EDEDED}[\overhang]
{\cellcolor{navy}\abinput{macro\_uj}} & \multicolumn{2}{c|}{1} & 2 & 3 & 4 \\ \hline

\rowcolor[HTML]{EDEDED}[\overhang]
{\cellcolor{navy}\abinput{nsppol}} & 1 & 2 & 2 & 2 & 2 \\ \hline

\rowcolor[HTML]{EDEDED}[\overhang]
{\cellcolor{navy}\textcolor{white}{Parameter}} & \multicolumn{2}{c|}{ \textcolor{copper}{Hubbard U} } &  &  & \textcolor{copper}{Hund's J} \\ \hline

\rowcolor[HTML]{EDEDED}[\overhang]
{\cellcolor{navy} \textcolor{white}{Perturbation applied to:}} & \multicolumn{2}{c|}{ $\alpha$ on both spin $\uparrow$ and spin $\downarrow$ } & $\alpha$ on spin $\uparrow$ & $\alpha$ on spin $\uparrow$ & $+\beta$ on spin $\uparrow$; $-\beta$ on spin $\downarrow$ \\ \hline

\rowcolor[HTML]{EDEDED}[\overhang]
{\cellcolor{navy} \textcolor{white}{Response monitored on:}} & \multicolumn{2}{c|}{ spin $\uparrow\ +$ spin $\downarrow$ } & spin $\uparrow$ & spin $\downarrow$ & spin $\uparrow\ -$ spin $\downarrow$ \\

\end{tabular}}
\caption{Variables \texttt{macro\_uj} and \texttt{nspden} combinations available in \Abinit\ and their corresponding Hubbard parameter.}
\label{macro_uj_options}
\end{table}

It is worth highlighting that the J calculated here using \abinput{macro\_uj}\texttt{=3} and \abinput{nsppol}\texttt{=2} is \emph{not} the Hund's J parameter. For the purposes of calculating U, we rely primarily on \abinput{macro\_uj}\texttt{=1} and \abinput{nsppol}\texttt{=2}. This setting will apply the same potential shift to both the up and down spin channels and monitor the occupancy response on the sum of occupancies on those same spin channels.

The strength of the perturbation is determined by \abinput{pawujv}. The default units for this variable are Hartree, but other units (notably eV) may also be specified. The variable \abinput{pawujat}, a single integer, specifies the atom number (the atom coordinate index listed under \abinput{xred} or \abinput{xcart}) on which the perturbation is to be applied. Make sure this is the same atom specified as a separate species in generating the \texttt{WFK} file in Step 1.

\begin{center}
\begin{tcolorbox}[width=0.8\linewidth,colback={lightgrey},title={NOTE: Reading in the right WFK file},colbacktitle=sea,coltitle=white]
    The default settings adopted when \abinput{macro\_uj} is non-zero are \abinput{tolvrs}\texttt{=1d-8} and \abinput{irdwfk}\texttt{=1}. The former dictates the tolerance on the potential residual (i.e., the difference between the input and output potentials pertaining to a particular SCF iteration). With the latter, \Abinit\ is instructed to read in the \texttt{WFK} files for a prior run, given the files are named according to a specific convention. Alternatively, we can specify the \texttt{WFK} file and its path by name. To do this, set the variable \abinput{getwfk\_filepath}\texttt{="</path2file/filename\_WFK>"}.
\end{tcolorbox}
\end{center}

The input parameter named \abinput{dmatpuopt}, of which there are four options, selects the expression with which the density matrix elements for each subspace are calculated using PAW projectors. These options are discussed in Section \ref{dmatpuopt}, and we refer the reader to reference \cite{MacEnulty2023} for a comprehensive evaluation of the influence of this variable on the Hubbard parameters.

\begin{figure}
    \centering
    \includegraphics[width=1.0\linewidth]{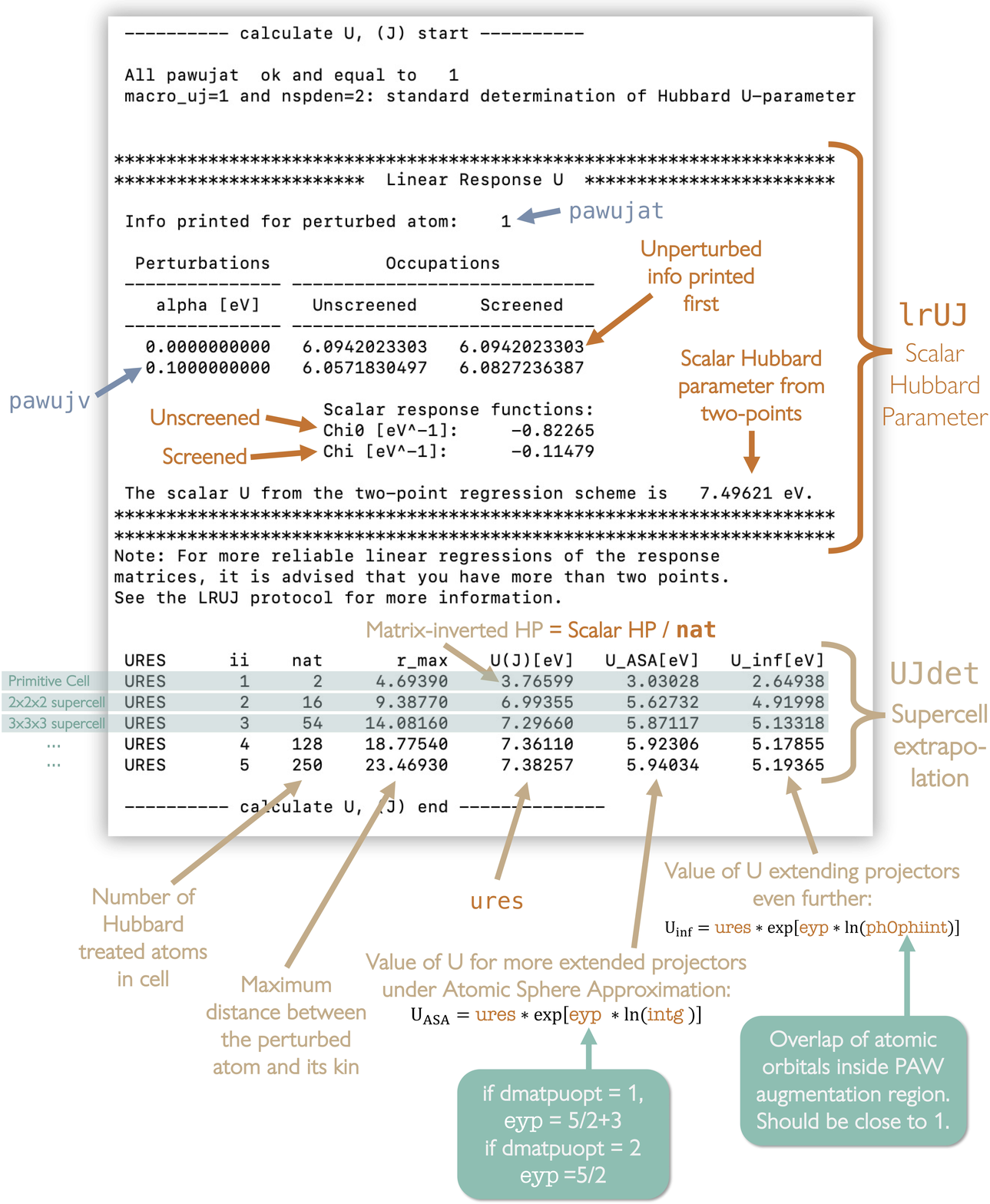}
    \caption{Generic output of the two-point Hubbard parameter determination functions, found in the \Abinit\ output file (typically with suffix \texttt{.abo}) for a linear response DFT run. Output for the Hubbard U is shown here. In the case of a Hund's J calculation (obtained with \abinput{macro\_uj}\texttt{=4}), the labels referencing \texttt{U}, \texttt{alpha}, and \texttt{Occupations} will be replaced by \texttt{J}, \texttt{beta}, and \texttt{Magnetizations}, respectively. The unscreened response \texttt{Chi0}, as printed, is treated with \abinput{diemix} (see appendix \ref{Mixing_Schemes} and section \ref{PlottingLR} for clarification). Explanation of the \ujdet\ supercell extrapolation output written to \texttt{.abo} file. See appendix \ref{ujdet_calculation} for more information regarding the intrinsic variables. Note that the small unit cell used here is only for illustrative purposes and will not yield converged (scalar or matrix) parameters.}
    \label{ujdet_output}
\end{figure}

Lastly, to have the \ujdet\ internal functions print out a verbose level of information as it completes its routine, the variable \abinput{pawprtvol} should be set to -3. To further manage the print volume, set \abinput{prtvol} as needed.

In changing only these variables, we set up only one perturbative calculation. This is sufficient to avail of the \ujdet\ utility functionalities, which require only two data points as discussed above. However, in many, if not all, cases, one perturbation is inadequate to compute a good regression of the linear response data, and no error analysis can be conducted thereof.

For this reason, we will need to conduct several (at minimum two, although the more, the better) perturbative calculations. We will take advantage of \Abinit’s dataset function to get our system to iteratively undergo $n$ perturbations by setting \abinput{ndtset} to $n$ and then specifying which perturbation strengths \abinput{pawujv1}, \abinput{pawujv2}, ... , \abinput{pawujv\emph{n}} we would like to apply. Once completed, launch the run.

Once the datasets have converged, your directory will have $n$ files with the suffix \texttt{LRUJ.nc}. These files, which are NetCDF binaries, contain all the internal information pertaining to the perturbations undergone. The \lruj\ utility will read in a series of these files and harvest the necessary information to calculate the selected Hubbard parameter.

\begin{center}
\begin{tcolorbox}[width=0.8\linewidth,colback={lightgrey},title={Developer's Note: Info in \texttt{LRUJ.nc} file},colbacktitle=sea,coltitle=white]
    For those considering development on the \lruj\ utility, the internal variables stored in the \texttt{LRUJ.nc} files are as follows: \abicode{nnat}, \abicode{natom}, \abicode{ndtpawuj}, \abicode{nspden}, \abicode{nsppol}, \abicode{usepaw}, \abicode{macro\_uj}, \abicode{pawujat}, \abicode{dmatpuopt}, \abicode{diemix}, \abicode{diemixmag}, \abicode{ph0phiint}, \abicode{uj\_pert}, \abicode{luocc}.
\end{tcolorbox}
\end{center}

In the \Abinit\ output file, all information related to the two-point calculation of the scalar Hubbard parameter and all information regarding the \ujdet\ functionalities (completed once for every dataset) can be found between the ``\texttt{calculate U, (J)}" flags. An annotated example of the standard, scalar Hubbard parameter output, in addition to the output of the supercell extrapolation scheme from \ujdet, comprises figure \ref{ujdet_output}.

\subsubsection{\label{Step3LRUJ}\textcolor{lightgray}{Execution of the \emph{\lruj}\ post-processing utility}}

Once the \texttt{LRUJ.nc} files are printed, execute the \lruj\ post-processing utility with the following command.

\begin{center}
\texttt{lruj *\_LRUJ.nc > lruj.out}
\end{center}

\begin{figure*}
    \begin{minipage}{\textwidth}
        \includegraphics[trim={0.5cm 0.35cm 0.8cm 0.35cm},clip,width=\textwidth]{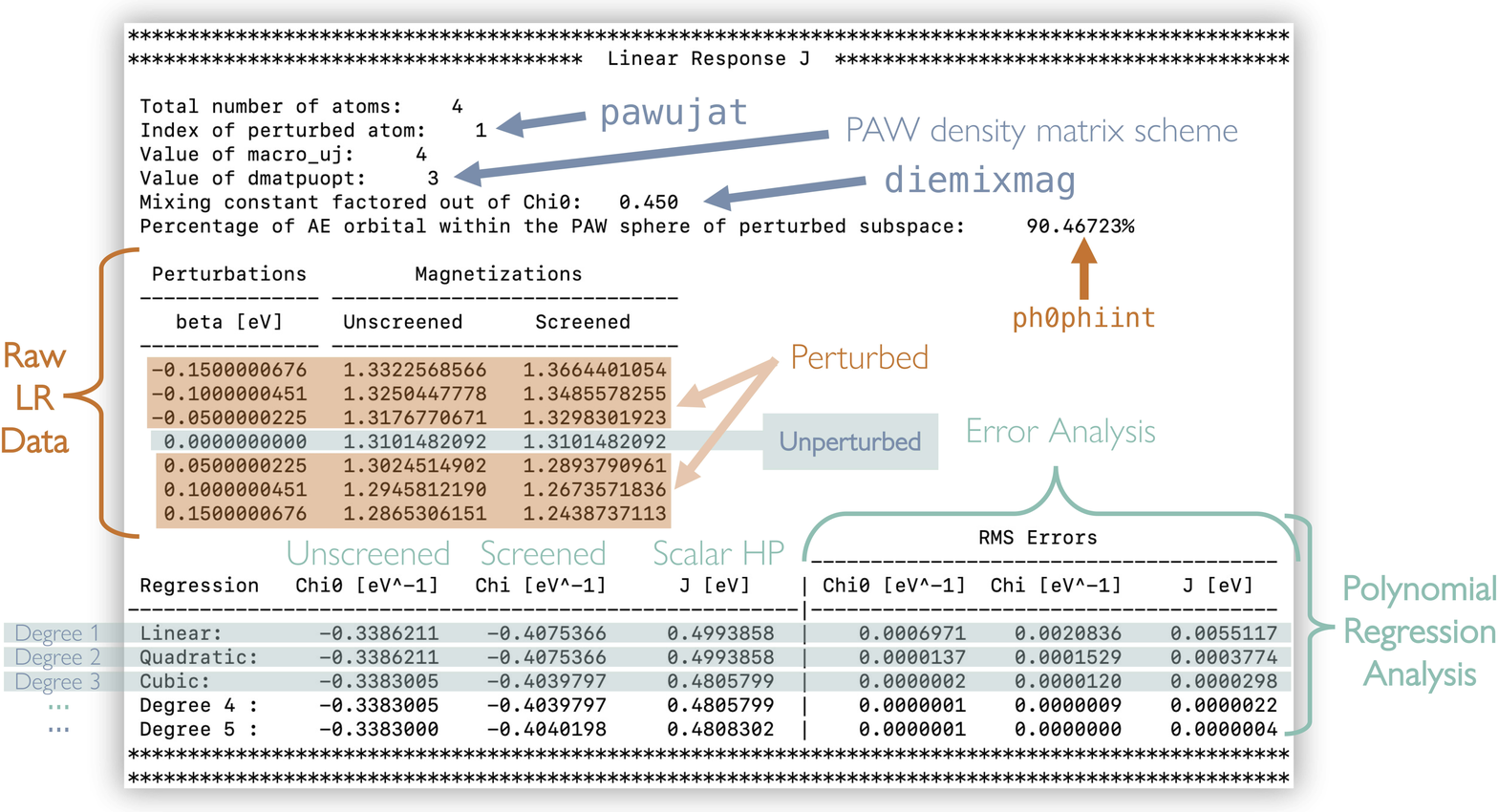}
        \caption{Output of the \lruj\ post-processing utility for the Hund's J parameter, having requested a maximum polynomial degree of 5 via command line. Note that this output is from Abinit Version 9.10.5 and so contains erroneous HP RMS errors in column 7. See Section \ref{LRUJpostproc} for more details.}
       \label{lruj_output}
    \end{minipage}
\end{figure*}

It should take less than a second to run. If the \lruj\ utility runs successfully, the resulting output file, \texttt{lruj.out}, should resemble that shown in figure \ref{lruj_output}. The calculation shown looks at the Hund's J parameter (\abinput{macro\_uj}\texttt{=4}) using results from 6 perturbations, the strengths of which are listed in the first table alongside the corresponding subspace magnetizations, both unscreened (for $\chi_{0_\textrm{M}}$) and screened (for $\chi_\textrm{M}$).

The last table gives the values for $\chi_0$ ($\chi_{0_\textrm{M}}$), $\chi$ ($\chi_\textrm{M}$), the Hubbard U (J), and their RMS errors in units of eV, for all polynomial regressions up to degree 3 (cubic), by default. One has the option to calculate higher-order polynomials, up to degree $n - 2$ for $n$ points. This is done by appending the degree option \texttt{--d <\textit{maximum degree}>} to the command line. For example, for the example calculation with 7 data points, one can bash

\begin{center}
    \texttt{lruj *LRUJ.nc --d 5 > lruj\_d5.out}
\end{center}

\noindent to get parameters and errors corresponding to all polynomials of order 1 through 5, as shown in figure \ref{lruj_output}. Other command line options for the \lruj\ utility include \texttt{--version} and \texttt{--help}.

The values in eV of the Hund’s J parameter according to each regression are found in column four. To assess which one is best, you’ll want to use the RMS errors in column seven (more information on how the error analysis is conducted in Section \ref{LRUJpostproc}) in addition to the visual behavior of the linear response, which can and should be plotted (see Section \ref{PlottingLR}), particularly if the RMS errors seem unusually large.

\begin{center}
\begin{tcolorbox}[width=0.8\linewidth,colback={lightgrey},title={NOTE: Polynomial order in \texttt{lrUJ}},colbacktitle=sea,coltitle=white]
   The \lruj\ utility is programmed to only calculate polynomials up to degree \textit{n – 2} for \textit{n} data points, a conservative measure implemented to avoid spurious overfitting effects. By default, then, if one reads in only two \texttt{LRUJ.nc} files, the maximum polynomial regression conducted will be linear; three \texttt{LRUJ.nc} files means maximum degree is quadratic, and so on, up to degree 3 (cubic). If higher polynomial order regressions are needed, and the number of data points suffices, then use the \texttt{--d <\textit{maximum degree}>} command line option to maximum degree.
\end{tcolorbox}
\end{center}

At the very end of the \lruj\ output file, information handy for plotting, such as the coefficients of the polynomial regression formulae, is printed in YAML format.

\subsubsection{\label{PlottingLR}\textcolor{lightgray}{Visualization of linear response data from \Abinit}}

Particular care must be taken when plotting the linear response data coming from \Abinit. The \lruj\ and \ujdet\ implementations both print out the raw data, meaning that the unscreened occupations (magnetizations) have not yet been scaled by the mixing constant \abinput{diemix} (\abinput{diemixmag}), as is necessary based on the conclusions of appendix \ref{Mixing_Schemes}. If one were to directly plot this raw data, the plot would show a slope that does not match the $\chi_0$ printed in the output file. To avoid this, we must perform a transformation on the data points, the form of which will be shown in the following proof. We assume a Hubbard U determination as an example proof, but the same deductions follow for the Hund's J parameter.

\begin{figure*}[t]
    \begin{minipage}{\textwidth}
        \includegraphics[trim={0.05cm 0.25cm 0.35cm 0.35cm},clip,width=\textwidth]{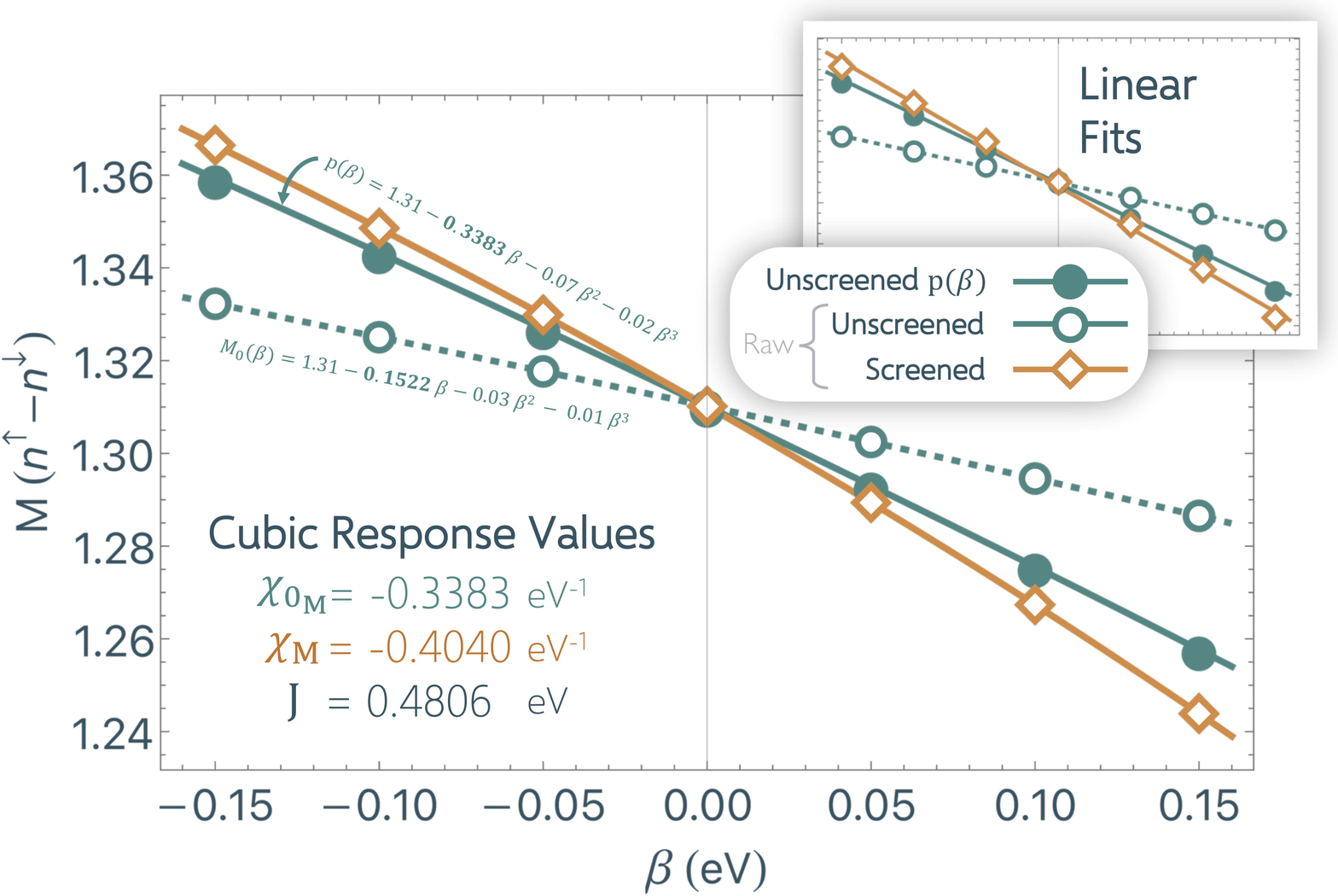}
        \caption{Linear response Hund's J plot demonstrating the necessary transformation of the raw \Abinit\ unscreened data to account for the mixing constant \abinput{diemixmag} (or \abinput{diemix} in the case of the Hubbard U). The example data used here is the same as that of figure \ref{lruj_output}, which is more aptly fitted with a degree 3 (cubic) polynomial as opposed to a linear function. The cubic fits are shown in the main figure, and the linear fits for the same data are shown in the upper right inset for qualitative comparison. The $p(\beta)$ function takes the same form as equation (\ref{PlottedUnscreenedResponse}), noting that we substitute $M_0$ for $N_0$ and $\beta$ for $\alpha$ to accommodate the Hund's J as opposed to the Hubbard U.}
       \label{PlottingTransform}
    \end{minipage}
\end{figure*}

We will refer to the data set of unscreened occupations as $N_0$, to which some polynomial function of the perturbation strength $\alpha$ is fitted, producing a regression function $N_0(\alpha)$. The unscreened response function is defined as

\begin{equation}\label{PlotProof1}
    \chi_{0}=\frac{1}{\theta}\left.\frac{dN_0}{d\alpha}\right|_{\alpha=0}
\end{equation}

\noindent where $\theta=$\abinput{diemix}. In order to plot the unscreened response data with a function $p(\alpha)$ set such that $\chi_{0}$ is shown with its $\theta$-corrected slope at the zero-perturbation axis, we perform the following multiplicative transformation on $N_0(\alpha)$,

\begin{equation}\label{PlotProof2}
    p(\alpha)=\gamma\cdot N_0(\alpha)\ +\ c
\end{equation}

where $\gamma$ and $c$ are constants. The mandatory criterion governing the shape of $p(\alpha)$ is

\begin{equation}\label{PlotProof3}
    \left.\frac{dp}{d\alpha}\right|_{\alpha=0}=\chi_{0}.
\end{equation}

It follows from equations (\ref{PlotProof1}), (\ref{PlotProof2}) and (\ref{PlotProof3}) that

\begin{align}
    \left.\frac{dp}{d\alpha}\right|_{\alpha=0}&=\gamma\left.\frac{dN_0}{d\alpha}\right|_{\alpha=0} \\
    &=\chi_{0} \\
    \Rightarrow\ \gamma\ \theta\ \chi_{0}&=\chi_{0} \\
    \therefore\gamma&=\frac{1}{\theta}
\end{align}

In order to find the second constant $c$, we impose a secondary criterion on the shape of $p(\alpha)$ to ensure that $p(0)=N_0(0)$. This leads to

\begin{align}
    p(0)&=\frac{N_0(0)}{\theta}\ +\ c=N_0(0) \\
    \therefore c&=\frac{N_0(0)\ \left(\theta-1\right)}{\theta}
\end{align}

Thus, 

\begin{equation}\label{PlottedUnscreenedResponse}
    p(\alpha)=\frac{1}{\theta}\ \left(N_0(\alpha)\ +N_0(0)\ \left(\theta-1\right)\right).
\end{equation}

As mentioned earlier, the same conclusion can be reached assuming a $\beta$ perturbation coupled with \abinput{diemixmag} as $\theta$ for the Hund's J parameter. Figure \ref{PlottingTransform} demonstrates such a transformation for the raw Hund's J \lruj\ data shown in figure \ref{lruj_output}. 


One can customize the mixing constant-corrected linear response plot by importing the perturbation/occupation table into one's choice graphical utility. Plot the screened occupations as they are printed; for the unscreened occupations, plot the function in equation (\ref{PlottedUnscreenedResponse}). Figure \ref{PlottingTransform}, for example, was generated with Mathematica.

\subsection{\label{InternalWorkings} Internal workings of the \Abinit\ \emph{U(J)} determination procedures}

In this section, we describe in detail the algorithm \Abinit\ undergoes to conduct linear response calculations. While sufficient as a launchpad for future developers of the program, this description is primarily intended to provide \Abinit\ users with a more transparent understanding of the linear response operations and their connection with the objects printed in the output files. We follow the internal variables as they undergo transformations and transfer relevant information, monitoring how the potential perturbations render occupancy responsesrender Hubbard parameters through the \lruj\ post-processor. For ease of reference, we use the language intrinsic to \Abinit, referencing variables, functions, subroutines, modules and programs as they are in the \Abinit\ source code.

\subsubsection{\label{perturb}\textcolor{lightgray}{Application of perturbation}}

Say we launch an \Abinit\ run in which we seek to determine a Hubbard parameter for subspace $n\ell_\textrm{U}$ by perturbing the potential of atom \abinput{pawujat} by a strength \abinput{pawujv}.

After the variables are read from the input file and the ground state driver is activated, the non-zero \abinput{macro\_uj} flag sets off the self-consistent cycle driven by \abicode{pawuj\_drive}. Here, the strength of the perturbation \abinput{pawujv} is read and stored in a matrix called \abicode{atvshift}—a \abinput{nsppol} $\times\ (2*\ell_\textrm{U}+1)$ array—according to the type of perturbation incited via the value of \abinput{macro\_uj}. (For example, for the Hubbard U on a $3d$ subspace, \abicode{atvshift} will be a $2\times5$ matrix in which all elements are equal to \texttt{+}\abinput{pawujv}. Alternatively, for the Hund's J parameter, the second row of \abicode{atvshift}, representing the spin-down channel, will be set equal to \texttt{-}\abinput{pawujv}, keeping the first row equal to \texttt{+}\abinput{pawujv}.) Following this, the PAW density is initialized and the unperturbed occupancy matrix calculated, diagonalized, and printed. Here is where the first linear response data point, corresponding to the unperturbed ground state read-in via the \texttt{WFK} file, is collected. The collection occurs in subroutine \abicode{pawuj\_red}. More information on how and what information is collected is described in Section \ref{delta_occupation}.

It is important to note that at this point in the code, the density (and thus the potential) is mixed according to the mixing scheme outlined in appendix \ref{Mixing_Schemes}. This means that the potential perturbation applied in the first iteration of the SCF cycle will be scaled by the value of \abinput{diemix}, which is equal to 0.45 by default when \abinput{macro\_uj}\texttt{>0}. We must accordingly descale the unscreened response function $\chi_0$ when the time comes.

The program then calls subroutine \abicode{pawdij}. This is where the program computes the pseudopotential strengths $D_{ij}$ of the non-local Hamiltonian operator. The potential

\begin{equation}\label{PAW_ham}
    H^{\textrm{PAW}}=-\frac{1}{2}\Delta+\tilde{\nu}_\mathrm{eff}+\sum_{ij}\left.\ |{\widetilde{p}}_i\right\rangle\left({\hat{D}}_{ij}+D_{ij}^1-{\widetilde{D}}_{ij}^1\right)\left\langle{\widetilde{p}}_j|\right.
\end{equation}

\noindent Here, -$\frac{1}{2}\Delta$ is the kinetic energy operator and $\tilde{\nu}_\mathrm{eff}$ is the effective one-electron potential written in the PAW formalism. As in Section \ref{PAW Formalism}, the indices $i,j$ refer to congruent but distinct sets of four indices: $\{t,\ell,m_{\ell},n\}$. The non-local part of the Hamiltonian mirrors the PAW energy, 

\begin{equation}
    E=\hat{E}+E^1-{\widetilde{E}}^1
\end{equation}

\noindent such that ${\hat{D}}_{ij}=\partial\hat{E}/\partial\rho_{ij}$ is the derivative of the PAW pseudized energy with respect to the density. Similarly, $D_{ij}^1=\partial E^1/\partial\rho_{ij}$ and $\tilde{D}_{ij}^1=\partial \tilde{E}^1/\partial\rho_{ij}$, where $E^1$ and $\tilde{E}^1$ are, respectively, the all-electron and pseudized on-site energies.

Implementing this formalism in \Abinit\ requires categorization of these terms into those calculated outside the SCF loop and those calculated within the SCF loop. For a more detailed derivation of this, see reference \cite{torrent_implementation_2008}. For now, and for our purposes, it suffices to note that the pseudopotential strengths $D_{ij}$ of the non-local Hamiltonian operator for each spin channel are calculated inside \Abinit\ as the following sum of terms,

\begin{equation}\label{pseudo_strength_total}
D_{ij}=D_{ij}^{0}+\hat{D}_{ij}+D_{ij}^\textrm{H}+D_{ij}^\textrm{xc}+D_{ij}^\textrm{U},
\end{equation}

\noindent where $D_{ij}^\textrm{U}$ is the term to which the $\alpha$ perturbation is applied, and $D_{ij}^{0}$, $D_{ij}^\textrm{H}$, and $D_{ij}^\textrm{xc}$ are, respectively, the atomic, Hartree, and exchange-correlation components, all of which are functionals of the density \cite{torrent_implementation_2008}. In the \Abinit\ source code, this matrix is named \abicode{dijpawu}.

Inside the \abicode{pawdij} driver, a loop over all atoms in the cell is induced, within which the subroutine \abicode{pawdiju} is summoned. Here, matrix \abicode{dijpawu} is defined as a function of the spin channel (either 1 or 2 for up or down, respectively) and of the matrix indices, which enumerate the non-core electrons by systematically combining the principle quantum number $n$, angular quantum number $l$, the magnetic quantum number $m_l$, the PAW projector index $n$, and the location in the matrix. In other words, the 2-dimensional matrix across all sub-indices pertaining to $i$ and $j$ is unfolded into a one-dimensional vector.

As an example of how this works, consider the case in which we apply a perturbation $\alpha$ to the spin-up $3d$ orbital of Ni. Say, we use a PAW dataset for Ni that has two partial waves to describe both the $p$ and $d$ orbitals, but only one plane wave for $s$ orbitals, and freezes a core containing all orbitals with $n\leq2$. For one spin channel on one atom, we are left with 18 combinations of quantum numbers $n>2$, $l$, $m_l$ and PAW projector $n$ (the $4s$ orbital of Ni contributing 2 electrons $\times$ 1 partial wave = 2 elements, and the $3d$ orbital contributing 8 electrons $\times$ 2 partial waves = 16 elements). So, the matrix \abicode{dijpawu} for each spin channel is $18 \times 18$, yielding 324 matrix elements. However, the pseudopotential strengths are symmetric across the diagonal (i.e., element $ij$ = element $ji$); to save memory and time, \Abinit\ computes the upper right triangular matrix elements only (a total of 171 in our Ni example). The matrix elements are thus enumerated from 1 to 171 starting from the upper left and reading left to right, top to bottom.

The $D_{ij}^\textrm{U}$ matrix elements themselves are found to be

\begin{equation}
\abicode{dijpawu}=\langle\phi_{n_i}|P_{mm'}^{t_i}|\phi_{n_j}\rangle\ast V_{m_im_j}^{\sigma,\textrm{U}}.
\end{equation}

Here, $P_{mm'}^{t_i}$ is the AE projection operator acting on the radial parts of the PAW AE basis functions $\phi_n$. This term is discussed in detail in Section \ref{dmatpuopt}. Furthermore, the term $V_{m_im_j}^{\sigma,\textrm{U}}$ comprises a homogeneous potential $V^{\sigma,\textrm{U}}$ across all $m_i$ for a particular subspace and, if appropriate, the perturbation:

\begin{align}
V_{m_im_j}^{\sigma,U}=
\begin{cases} 
    V^{\sigma,\textrm{U}} & m_i\neq m_j \\
    V^{\sigma,\textrm{U}}+\abicode{fatvshift}*\abicode{atvshift}(\sigma,m_i) & m_i=m_j.
\end{cases}
\end{align}

The term \abicode{fatvshift} is vestigial from prior versions of the \ujdet\ implementation, where a loop over values \abicode{fatvshift}\texttt{=1} and \abicode{fatvshift}\texttt{=-1} corresponded to the positively and negatively valued perturbations of strength \abinput{pawujv}. (Now, \abicode{fatvshift}\texttt{=1} only). If \abinput{pawprtvol}\texttt{=3} in the input file, the entire \abicode{dijpawu} matrix will be printed in the \texttt{.log} file for all spin channels and all atoms, where one can verify that the perturbation is, indeed, being applied here. Here ends the subroutine \abicode{pawdiju}, which returns \abicode{dijpawu} for each spin channel. Back in \abicode{pawdij}, \abicode{dijpawu} is added via matrix addition to all other pseudopotential strength matrices to calculate the total $D_{ij}$ in accordance with equation (\ref{pseudo_strength_total}).

\subsubsection{\label{dmatpuopt}\textcolor{lightgray}{Calculation of orbital occupancies via \emph{\abinput{dmatpuopt}}}}

Calculation of the occupancy matrix, or more precisely choice of the projection operator, is dictated by the input variable \abinput{dmatpuopt}, which may take on values one through four. More information on this topic can be found in references \cite{Amadon2008,MacEnulty2023,Geneste2017}.  Briefly, subspace occupancies in the PAW formalism may be calculated using the AE projection operator $P_{mm'}^{t_i}$ via

\begin{equation}
    n_{mm'}^{t_i\sigma}=\sum_{ij}{\rho_{{\sss ij}}^\sigma\langle\phi_{n_i}|P_{mm'}^{t_i}|\phi_{n_j}\rangle},
\end{equation}
where $\phi_n$ are the radial parts of the PAW AE basis functions, and the density matrix inside the PAW augmentation region is
\begin{equation}
    \rho_{{\sss ij}}^\sigma=\sum_{k,v}{f_{kv}^\sigma \langle {\widetilde{\Psi}}_{kv}^{\sigma} |{\widetilde{p}}_{\sss i}\rangle\langle{\widetilde{p}}_{\sss j}| {\widetilde{\Psi}}_{kv}^{\sigma} \rangle}.
\end{equation}

When \abinput{dmatpuopt}\texttt{=1}, occupations are projections on bound state atomic orbitals $\phi_0$,

\begin{equation}\label{nocc_project}
    n_{m,m\prime}^{t_i,\sigma}=\sum_{ij}{\rho_{{\sss ij}}^\sigma\left\langle\phi_{n_i}\middle|\phi_{0}\right\rangle\left\langle\phi_{0}\middle|\phi_{n_j}\right\rangle}.
\end{equation}

The \Abinit\ documentation for this variable is clear that the \abinput{dmatpuopt}\texttt{=1} option must be accompanied by a PAW dataset wherein the first atomic wavefunction of the correlated subspace (that which is set to $\phi_0$ in \Abinit) is a normalized atomic eigenfunction. To determine if a particular PAW dataset meets this criterion, one must refer to the documentation of its generator.

We are able to reasonably infer, based on the \href{https://users.wfu.edu/natalie/papers/pwpaw/atompaw-usersguide-MarcTorrent.pdf}{\atompaw\ user guide} in addition to reference \cite{holzwarth_projector_2001}, that PAW datasets generated by \atompaw\ will always feature a normalized atomic eigenfunction as the first atomic wavefunction of an atomic dataset. Step 4 on Page 2 of reference \cite{holzwarth_projector_2001} states clearly that \atompaw\ mandates the use of ``atomic eigenfunctions related to valence electrons (bound states)" as the partial waves included in the PAW basis. Therefore, all PAW datasets generated by \atompaw, including the JTH sets listed on PseudoDojo \cite{pseudodojo2018}, list atomic eigenfunctions as the first atomic wavefunctions of the correlated subspace. The normalization, however, depends on the pseudo partial wave generation scheme. \atompaw\ provides two options for this scheme: the Vanderbilt or the Blöchl. Based on the descriptions of these schemes in Sections 1.1 and 1.2 of reference \cite{holzwarth_notes_nodate} and an \href{https://forum.abinit.org/viewtopic.php?f=9&t=3335&p=10247&hilit=dmatpuopt#p10247}{\Abinit\ forum response} in 2016, normalization of the pseudized basis functions and their corresponding projectors is guaranteed only under the Vanderbilt scheme. The JTH table of PAW datasets, available on the PseudoDojo website, therefore matches all criteria as a suitable dataset with which one may use \abinput{dmatpuopt}\texttt{=1}.

Because the PAW datasets most readily available for widespread use do not necessarily fulfill these criterion, \abinput{dmatpuopt}\texttt{=2} is established as the default setting. With \abinput{dmatpuopt}\texttt{=2}, occupations are proportional to projections of atomic orbitals onto each other,

\begin{equation}\label{nocc_no_project}
    n_{m,m\prime}^{t_i,\sigma}=\sum_{ij}{\rho_{{\sss ij}}^\sigma\left\langle\phi_{n_i}\middle|\phi_{n_j}\right\rangle}.
\end{equation}

Equation (\ref{nocc_no_project}) corresponds to a projection operator of form

\begin{equation}\label{dmat2_projector}
    P_{mm'}^{t_i}(\textbf{r},\textbf{r}') = 1_\circ(\textbf{r})\delta(|\textbf{r}'-\textbf{R}_\textbf{i}|-|\textbf{r}-\textbf{R}_\textbf{i}|)\times {Y}_{\ell m}(\hat{\textbf{r}})Y_{\ell m'}^\ast({\hat{\textbf{r}}}'), 
    \nonumber
\end{equation}

\noindent where $\delta$ is the Dirac-Delta function that effectively ``counts" spatial overlap, $1_\circ\left(\textbf{r}\right)$ is a step function equal to unity when $\textbf{r}$ is inside the augmentation region and zero elsewhere, and ${Y}_{\ell m}$ are the spherical harmonics. 


When \abinput{dmatpuopt}\texttt{=3} or \texttt{4},

\begin{equation}\label{nocc_dmat3}
    n_{m,m\prime}^{t_i,\sigma}=\mathcal{N}_{0}^{(2-\abinput{dmatpuopt})}\sum_{ij}{\rho_{{\sss ij}}^\sigma\langle\phi_{n_i}|\phi_{0}\rangle\langle\phi_{0}|\phi_{n_j}\rangle},
\end{equation}

\noindent where $\mathcal{N}$ is a normalization constant representing the overlap between the bound state atomic eigenfunctions inside the augmentation sphere, delimited by cutoff radius $r_\mathrm{c}$,

\begin{equation}\label{dmat_norm}
    \mathcal{N}=\int_0^{r_\mathrm{c}}{\phi_0^2\ dr}.
\end{equation}

The value is computed in subroutine \abicode{pawpuxinit} and printed in the log file as \abicode{ph0phiint(1)}. When \abinput{dmatpuopt}\texttt{=4}, $\mathcal{N}$ is squared in the denominator.

An evaluation of the effect of the choice of \abinput{dmatpuopt} on the magnitude of the Hubbard parameters can be found in reference \cite{MacEnulty2023}.

\subsubsection{\label{delta_occupation}\textcolor{lightgray}{Extraction of changes in occupation matrix}}

The outer SCF loop, declared in \abicode{pawuj\_drive}, continues after the perturbation is applied; the loop symmetrizes and prints $D_{ij}$. If \abinput{macro\_uj}\texttt{$>$0}, a subroutine labeled \abicode{pawuj\_red} is called. The subroutine generates the mesh that directly associates the strength of the perturbation (translated from \abicode{atvshift} to a shorter variable called \abicode{vsh}), and the corresponding change in occupancy, called \abicode{occ}. This information, along with the atom and spin indices, are saved in a type called \abicode{dtpawuj}, which is made accessible to the internal \ujdet\ and \lruj\ functions after the SCF cycle. The occupancy is calculated as the trace of the occupancy matrix $n_{m,m'}^{t_i,\sigma}$ discussed in Section \ref{dmatpuopt},

\begin{equation}
    \abicode{occ}(t_i,\sigma)=\text{Tr}[n_{m,m'}^{t_i,\sigma}]*(3-\abinput{nspden}).
\end{equation}

The factor $(3-\abinput{nspden})$ accounts for the occupation of two spin channels on a single atom if and only if we do not distinguish between the up and down spin channels. See table \ref{macro_uj_options} for clarity.

The SCF iteration concludes by calling the subroutines associated with updating the density and the Hamiltonian. It then uses those updated quantities to find the updated potential, restarts a new SCF iteration, and so on and so forth until self-consistency is achieved.

The combinatory values of \abinput{macro\_uj} and \abinput{nsppol} as outlined in table \ref{macro_uj_options} determine how the elements of \abicode{occ} are combined and saved in a new array called \abicode{luocc}. If \abinput{nsppol}\texttt{=1}, then \abicode{occ} and \abicode{luocc} are identical, complementing the application of the perturbation to the entire atomic subspace by monitoring the response on the entire atomic subspace. In the case of the Hubbard U calculation, however, where \abinput{nsppol}\texttt{=2} and \abinput{macro\_uj}\texttt{=1}, \abicode{luocc}($t_i$) $=$ \abicode{occ}($t_i,\uparrow$) $+$ \abicode{occ}($t_i,\downarrow$). In this way, the response is monitored on the total occupancy of the subspace. By contrast, when calculating the Hund's J by setting \abinput{nsppol}\texttt{=2} and \abinput{macro\_uj}\texttt{=4}, one monitors the subspace magnetization: \abicode{luocc}($t_i$) $=$ \abicode{occ}($t_i,\uparrow$) $-$ \abicode{occ}($t_i,\downarrow$).

The type \abicode{dtpawuj} saves four (\abicode{vsh}, \abicode{luocc}) pairs, indexed by integers 1-4. (In the verbose \texttt{.log} file, these pairs are labeled (\abicode{vsh1}, \abicode{occ1}), (\abicode{vsh2}, \abicode{occ2}), etc.; but the \abicode{occ} printed is actually the \abicode{luocc} value.) If the pair's referential index (called \abicode{iuj}) is an odd integer, that pair's occupation is harvested at the end of the first SCF cycle, immediately after the perturbation is applied to $D_{ij}$, but before the Hamiltonian and the density are updated to reflect that perturbation. These points will be used by \lruj\ and \ujdet\ to calculate the unscreened response matrix $\chi_0$. Conversely, if \abicode{iuj} is an even integer, that occupation is harvested after self-consistency has been achieved. Following suit, these points will be used in by the \ujdet\ and \lruj\ functions to calculate the screened response matrix $\chi$.

\begin{center}
\begin{tcolorbox}[width=0.8\linewidth,colback={lightgrey},title={Note: The unperturbed case},colbacktitle=sea,coltitle=white]
    The pairs (\abicode{vsh1}, \abicode{occ1}) and (\abicode{vsh2}, \abicode{occ2}) correspond to the unperturbed ground state. Because no perturbation is applied, the unscreened and screened responses are identical. That is, \abicode{vsh1}=\abicode{vsh2}=0.0 eV, and \abicode{occ1}=\abicode{occ2}, which are the ground state occupancies of the subspace in question.
\end{tcolorbox}
\end{center}

All (\abicode{vsh}, \abicode{luocc}) pairs for all atoms and spin channels are printed out at the end of the SCF iteration in which they are determined.

\subsubsection{\label{LRUJpostproc}\textcolor{lightgray}{The \emph{\lruj}\ post-processor}}

When the \lruj\ post-processing utility is executed via command line, it reads in a user-specified series of NetCDF files with suffix \texttt{LRUJ.nc}. The program sorts the $n$ files in order of perturbation strength, then reads in all necessary data related to those perturbations, including the unperturbed state (which is output in all perturbative calculations) in addition to the unscreened and screened occupations (or spin magnetization in the case of Hund’s J). 

\begin{center}
\begin{tcolorbox}[width=0.8\linewidth,colback={lightgrey},title={NOTE: units},colbacktitle=sea,coltitle=white]
    After the perturbation strengths are read from the NetCDF files, where they are reported in Ha, the \lruj\ program converts them to eV. All calculations are then conducted in eV. 
\end{tcolorbox}
\end{center}

Compatibility tests are conducted on the input information, and then the linear response procedure begins.

\begin{figure*}[t]
    \begin{minipage}{\textwidth}
        \centering
        \includegraphics[width=1.0\textwidth]{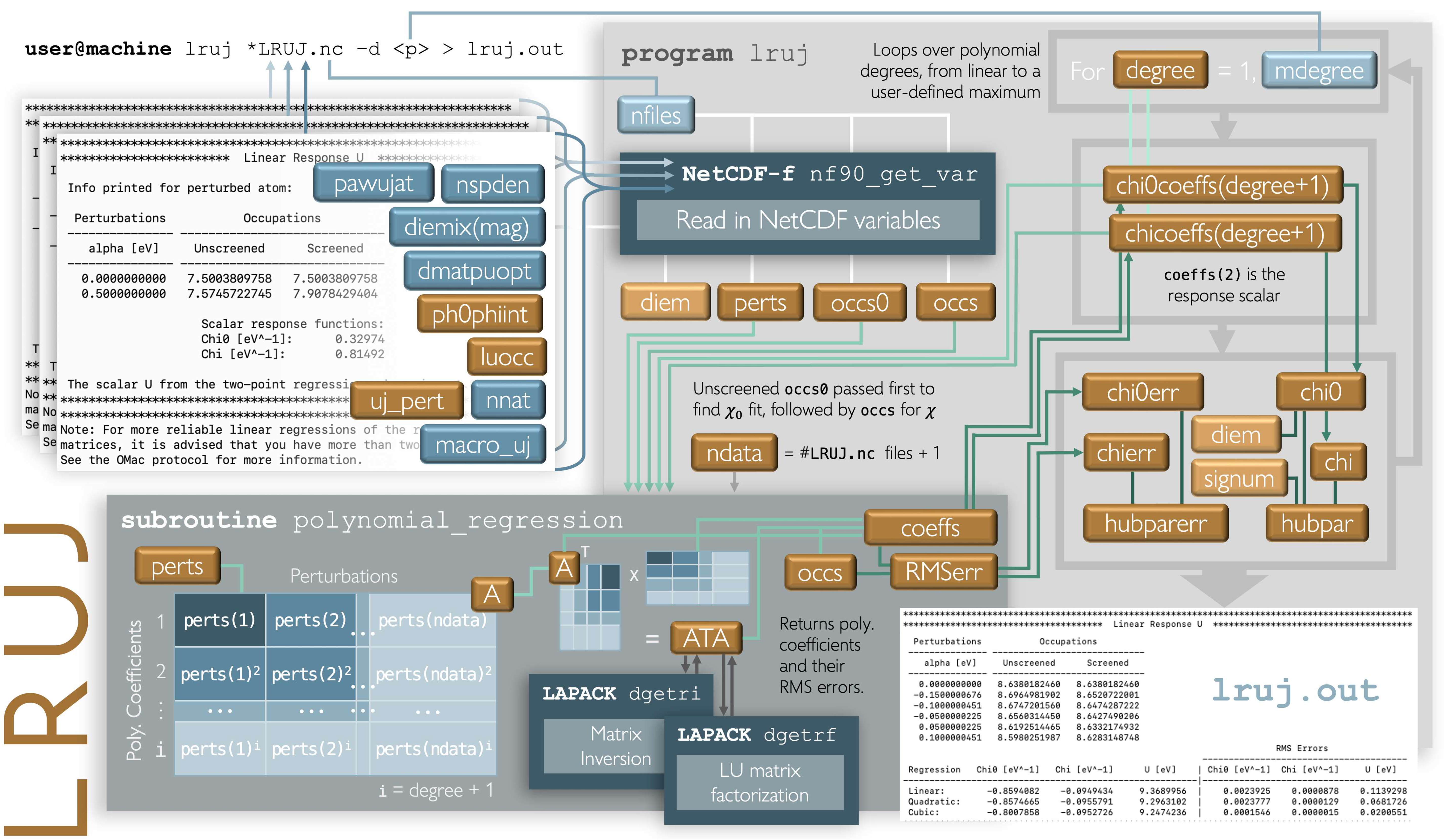}
        \caption{Flowchart illustrating the \lruj\ algorithm.}
       \label{lruj_flowchart}
    \end{minipage}
\end{figure*}

The \lruj\ utility was constructed with the occasional non-linearity of linear response in mind. Therefore, the program has an inbuilt subroutine that calculates the polynomial regression of any degree for any list of data points. This subroutine calculates the $n+1$ coefficients of a degree $n$ polynomial by constructing a matrix using the fitted data points, then solving the resulting system with linear algebra. The mathematical specifics of this procedure are illustrated in figure \ref{lruj_flowchart} but outlined more formally \href{https://muthu.co/maths-behind-polynomial-regression/#:~:text=Polynomial%20regression%20is%20a%20process,is%20a%20set%20of%20coefficients.}{on this website}. Thus, as shown in figure \ref{lruj_flowchart}, the \abicode{polynomial\_regression} subroutine is dependent on the functionalities available in LAPACK.

We use polynomials only in this program because of their relative simplicity and reasonably predictable RMS error behavior. In case the user wants to fit another type of function to the data, the data points are printed out in an easily copy-pasted or parsed format for independent regression analysis.

To begin the regression procedure, the \lruj\ program tests if the user has specified a maximum polynomial degree to calculate. If so, this degree has to be greater than or equal to the number of data points (i.e., the number of incoming \texttt{LRUJ.nc} files plus one unperturbed state) plus 2. If the user has left this information unspecified, then the maximum polynomial degree will default to cubic (degree 3) UNLESS the number of input files is equal to two or three, in which cases the maximum polynomial degree will be set to linear (degree 1) or quadratic (degree 2), respectively.

Once the max polynomial degree is set, the arrays storing the response information for each degree are allocated and the loop over polynomial degree begins. For every degree, the polynomial regression subroutine is called twice: once to fit the unscreened occupancies (magnetizations) and record its unbiased RMS fit error, and the other to fit the screened occupancies (magnetizations) and record its unbiased RMS fit error. For an $N$-point linear regression $f\left(\alpha\right)$, where $\alpha_i$ is the perturbation strength corresponding to occupation (magnetization) $\mu_i$, the equation for unbiased RMS fit error is

\begin{equation}
\sigma_\textrm{RMS}=\sqrt{\frac{\sum_{i=1}^{N}\left(f\left(\alpha_i\right)-\mu_i\right)^2}{N-1}}
\end{equation}

The RMS fit error, alongside the fit coefficients, are returned to the main program, where the utility uses that information to find $\chi$ and $\chi_0$ as

\begin{equation}
\chi_0=\frac{1}{\abinput{diemix}}\left.\frac{df_0\left(\alpha\right)}{d\alpha}\right|_{\alpha=0}
\end{equation}

\begin{equation}
\chi=\left.\frac{df\left(\alpha\right)}{d\alpha}\right|_{\alpha=0}.
\end{equation}

With one-dimensional polynomials as functions of $\alpha$, the derivative at $\alpha=0.0$ eV is simply the second coefficient pertaining to that polynomial function. The unscreened response $\chi_0$ must be divided by the mixing parameter that was used in the preceding \Abinit\ run. This default mixing parameter is \abinput{diemix} and it is equal to, by default, 0.45. However, if the value of \abinput{diemix} is changed, or if a Hund’s J calculation is conducted (at which point \abinput{diemixmag} instead of \abinput{diemix} is used for the mixing constant), $\chi_0$ is divided by that value to get the true unscreened response.

The resulting scalar Hubbard parameter corresponding to these response functions is calculated as

\begin{equation}
\textrm{HP}=\abicode{signum}\cdot\left({{\ \chi}_0}^{-1}-\chi^{-1}\right)
\end{equation}

\noindent where \abicode{signum} = 1 if calculating the Hubbard U or \abicode{signum} = -1 if calculating the Hund’s J. The error on the Hubbard parameter, printed in column 7 of the \lruj\ output file, is then

\begin{equation}\label{rmsHP}
\sigma_\textrm{HP}=\sqrt{ \left( \frac{\sigma_{\chi_0}}{\chi_0^2} \right)^2 + \left( \frac{\sigma_\chi}{\chi^2} \right)^2}\ \ .
\end{equation}

Having executed its main function, the program concludes its operations by printing out the information in user-friendly format to the main output file (if specified in the command line; prints to terminal otherwise). An example of such an output is available in figure \ref{lruj_output}.

\begin{center}
\begin{tcolorbox}[width=0.8\linewidth,colback={lightgrey},title={NOTE: Initial bug fixes},colbacktitle=sea,coltitle=white]
    Although the \lruj\ post-processor emerged with \Abinit\ version 9.10.1, important bug fixes for this utility were implemented in \Abinit\ version 9.10.5, and a further update using the correct form of equation \ref{rmsHP} will go into effect in Abinit version 9.11. If using prior versions of \lruj, please note that the reported RMS error for the Hubbard parameter is unreliable (i.e., these versions of \lruj\ program erroneously do not square the response functions in the denominators of equation \ref{rmsHP}).
\end{tcolorbox}
\end{center}

\section{\label{Acknowledgements}Acknowledgements}

\blackout{LM acknowledges the Trinity College Dublin Provost PhD Project Awards. Calculations were principally performed on the Boyle cluster
at Trinity College Dublin, which was funded through grants from the European Research Council and Science Foundation Ireland and is maintained by Trinity Research IT.}

\newpage
\appendix
\addcontentsline{toc}{section}{Appendix}

\section*{\label{Appendix} Appendix}

\addcontentsline{toc}{section}{\spc\spc\spc\spc A: The PAW pseudopotential}
\subsection*{\label{PAW_pseudopotentials} Appendix A: The PAW pseudopotential}

PAW dataset generators construct a local ionic pseudopotential using the chosen PAW basis functions via a method that is closely analogous to Vanderbilt's ultrasoft pseudopotential generation, described in reference \cite{vanderbilt_soft_1990}. We briefly describe the construction of a PAW pseudopotential here from the basis functions defined earlier \cite{hine2017,kresse_ultrasoft_1999}.

The first step is to construct a screened local atomic pseudopotential $\tilde{\nu}_\mathrm{loc}$ for an atom in some reference configuration (usually the isolated atom), which is to be equivalent to the AE atomic potential $\nu_\mathrm{loc}$ beyond some radius $r_\mathrm{pp}$. (The cutoff radius $r_\mathrm{c}$ is not necessarily the same as $r_\mathrm{pp}$.) As an example of such a construction, Vanderbilt proposed the use of a zero-order spherical Bessel function (equation 58 in reference \cite{vanderbilt_soft_1990}).

The pseudopotential $\tilde{\nu}_\mathrm{loc}$ comprises contributions from both the core and valence densities. To expand the contribution of the latter, we take the pseudo partial waves—which have been defined such that outside of the cutoff radius $r_\mathrm{c}$ they are equivalent to the AE basis functions—and use them to generate a different set of functions $\chi_i$

\begin{equation}
    | \chi_i \rangle =(\epsilon_i-T-\tilde{\nu}_\mathrm{loc}(r)) | \tilde{\phi}_i\rangle
\end{equation}

\noindent where $\epsilon_i$ are the orbital energy eigenvalues associated with the AE basis functions and $T$ is the kinetic energy operator. The projectors $\tilde{p}_i$ are then represented in terms of $\chi_i$,

\begin{equation}
    | \tilde{p}_i \rangle =\sum_j P_{ij}^{-1} | \chi_i \rangle ,
\end{equation}

\noindent whereupon, owing to the orthogonality condition $\langle \tilde{p}_i | \tilde{\phi}_i \rangle=\delta_{ij}$,

\begin{equation}
    P_{ij} = \langle \tilde{\phi}_i | \chi_i \rangle.
\end{equation}

We take the operator $P_{ij}$ and employ it in the definition of a non-local potential $D_{ij}=P_{ij}+\epsilon_j Q_{ij}$, where $Q_{ij}$ is as defined in equation \ref{Qij}. In a final step, we effectively unscreen the potentials $D_{ij}$ and $\tilde{\nu}_\mathrm{loc}$ to deduce a valence potential

\begin{small}
\begin{equation}
    D_{ij}^0=P_{ij}+\epsilon_j Q_{ij}- \int{\tilde{\nu}_\mathrm{loc}\left(r\right)\left(\phi_i^\ast\left(\mathbf{r}\right)\phi_j\left(\mathbf{r}\right)-{\widetilde{\phi}}_i^\ast\left(\mathbf{r}\right){\widetilde{\phi}}_j\left(\mathbf{r}\right)\right)d\mathbf{r}}
\end{equation}
\end{small}

\noindent and a local ionic pseudopotential

\begin{equation}
    \tilde{\nu}_\textrm{H}[\tilde{n}_{Zc}] = \tilde{\nu}_\textrm{loc}-\nu_\textrm{H}[\tilde{n}^1+\hat{n}]-\nu_\textrm{xc}[\tilde{n}^1+\hat{n}+\tilde{n}_c],
\end{equation}

\noindent where $\nu_\mathrm{H}$ and $\nu_\mathrm{xc}$ are the Hartree and exchange-correlation potentials, respectively. Finally, the PAW pseudopotential representing the atom in its entirety is

\begin{equation}
    V_\mathrm{PP}=\tilde{\nu}_\textrm{H}[\tilde{n}_{Zc}]+\sum_{ij}\ D_{ij}^0|\widetilde{p_i}\rangle \langle\widetilde{p_j}|.
\end{equation}

\addcontentsline{toc}{section}{\spc\spc\spc\spc B: Mixing Schemes}
\subsection*{\label{Mixing_Schemes} Appendix B: Mixing Schemes}

\Abinit\ provides two mixing schemes: one that mixes the potential and one that mixes the density. Both are available in the PAW implementation, and both prove to perform equally well in efficiently achieving self-consistency (density mixing slightly outperforms potential mixing). However, density mixing is preferable when using PAW because of the degrees of freedom added to the electronic density via the pseudovalence density and the compensation charge density, the latter of which is directly related to the PAW occupation matrix. From reference \cite{gonze_abinit_2009}: \\

``When potential mixing is activated, all parts of the total energy are computed at the same time; the total energy is thus variational with respect to the self-consistent cycle step. When density mixing is activated, parts of total energy are computed at various stages of the cycle which results in a behavior of total energy that is not variational." \\

The new density is computed, mixed with previous densities, then used to update the energy total alongside other contributions that are not all updated at the same place in the SCF cycle. Inside PAW, the on-site density matrix $\rho_{ij}$, defined explicitly in reference \cite{torrent_implementation_2008}, is updated at the same level as the electronic density $\tilde{n}+\hat{n}$, and is then mixed at that level. Therefore, the $D_{ij}$ which is a potential term in PAW, is left unmixed by default \cite{torrent_implementation_2008}.

Density mixing is the default for \Abinit\ under the PAW protocol (\abinput{iscf}\texttt{ = 17}), specifically via the Pulay mixing algorithm \cite{pulay1980}, which was developed in 1980 as an efficient method of accelerating convergence of iterative sequences. Pulay mixing is used to mix $\rho_j^\mathrm{OUT}$ and the residual density $\rho_{j}^\mathrm{OUT}-\rho{j}^\mathrm{IN}$ in the following iterative update of the density,

\begin{equation}
    \rho_{i+1}^\mathrm{IN}=\rho_{i}+\mix[\rho_j^\mathrm{OUT},P(k)*(\rho_{j}^\mathrm{OUT}-\rho{j}^\mathrm{IN})],
\end{equation}

\noindent where $P(k)$ is a preconditioning factor corresponding to wavevector $k$, applied to the residual density $\rho_{j}^\mathrm{OUT}-\rho{j}^\mathrm{IN}$ of the prior iterations. This preconditioning factor is defined, by default, as the inverse of the model dielectric matrix

\begin{equation}\label{mixing_dielectric_function}
    P(k)=\epsilon^{-1}(k)=\abinput{diemix}* \frac{\abinput{diemac}^{-1}+(\abinput{dielng}*k)^2}{1+(\abinput{dielng}*k)^2}
\end{equation}

\noindent where \abinput{diemix} is the dielectric mixing constant, set to 0.7 by default for PAW calculations and 0.45 for linear response calculations; \abinput{diemac} is the model dielectric macroscopic mixing constant, which is typically very large for metals and around 10 for insulators. (\abinput{dielng} is a fine-tuning parameter). The variable \abinput{iprcell} can select the function used to define the preconditioning factor.

The dielectric mixing constant \abinput{diemix}, and its magnetic analog \abinput{diemixmag}, is applied to the first SCF density after the $\alpha$ ($\beta$) perturbation is applied but before the on-site orbital occupations (magnetizations) are calculated. This means that \abinput{diemix} (\abinput{diemixmag}) inadvertently scales the potential perturbation of the unscreened response matrix $\chi_0$ ($\chi_{0_\textrm{M}}$) in the determination of the Hubbard U (Hund's J) parameter. To counteract this, therefore, we must use the value of \abinput{diemix} (\abinput{diemixmag}) to unscale $\chi_0$ ($\chi_{0_\textrm{M}}$) in the Hubbard U (Hund's J) data-processing step. Based on a series of tests, we can say conclusively that changing \abinput{diemixmag} does not influence the Hubbard U parameter, and analogously, changing \abinput{diemix} does not influence the Hund's J.

\addcontentsline{toc}{section}{\spc\spc\spc\spc C: \ujdet\ prior to \Abinit\ version 9.9}
\subsection*{\label{AppendixSec:ujdet} Appendix C: \ujdet\ prior to \Abinit\ version 9.9}

\begin{figure}[b!]
    \centering
    \includegraphics[trim={0 0 0 0.25cm},clip,width=1.0\linewidth]{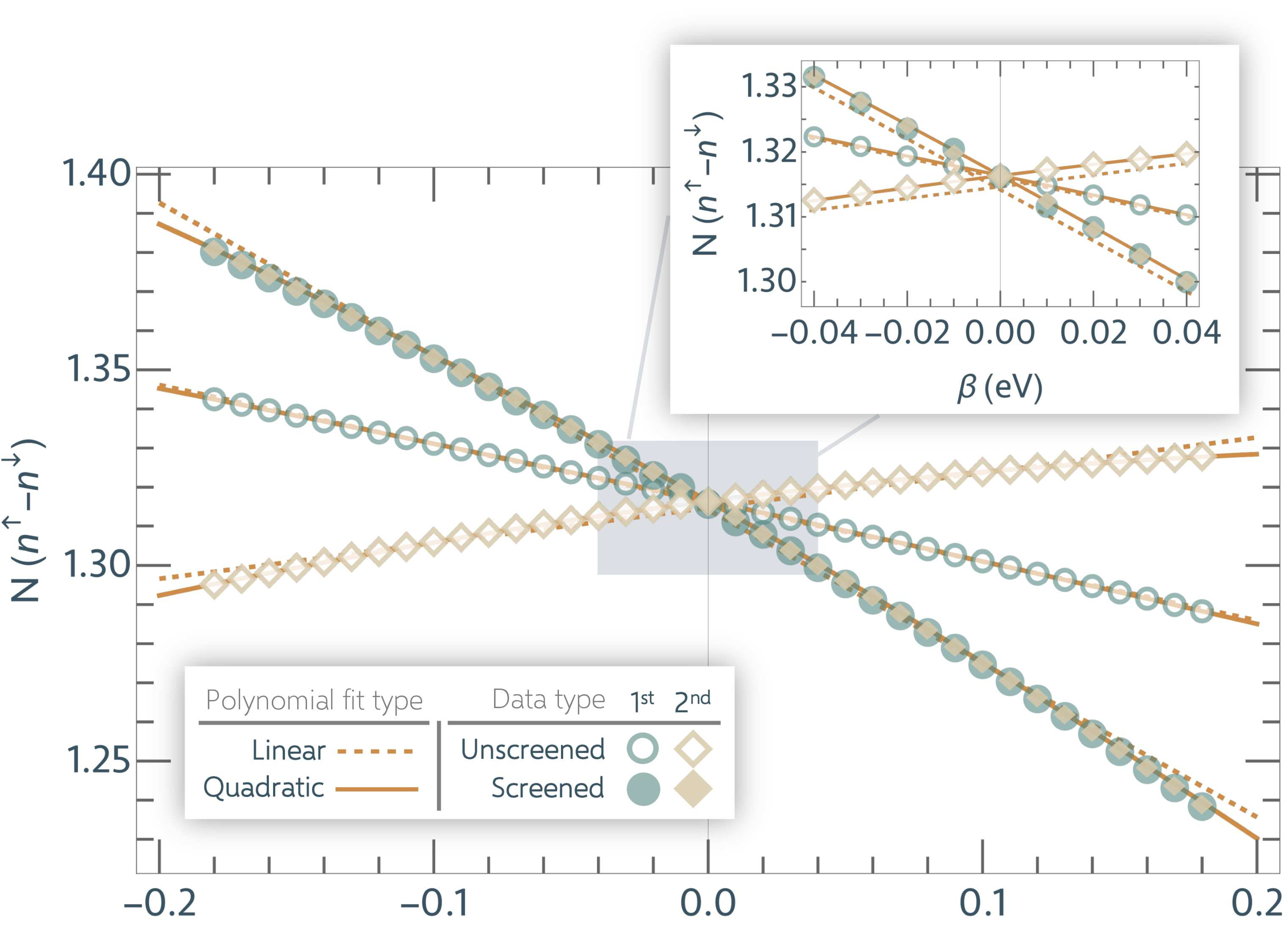}
    \caption[Caption for LOF]{Example of the linear response procedure conducted using the now outdated rendition of \ujdet\ in \Abinit\ version 9.6.2, to find Hund's J for Ni \textit{3d} orbitals on ferromagnetically ordered NiO \cite{footnote1}. Data points are raw perturbation-occupancy pairs output by \Abinit\ categorized according to their place in the calculation queue; that is, the SCF calculation involving the first ($\mathrm{1}^\mathrm{st}$) $\beta$ perturbation is represented by green circles, and the SCF calculation of the second ($\mathrm{2}^\mathrm{nd}$) $\beta$ perturbation is represented by tan diamonds. Open markers indicate unscreened response and filled markers indicate screened response. The screened response $\chi$ relaxes to the same value regardless of its status as first or second calculation. The unscreened responses, however, are different depending on which perturbation is applied first. This contributes to J parameters differing, in this case, by several eV. Orange lines are polynomial regressions (dashed for first-order polynomial, solid for second-order polynomial) of each set of data.}
    \label{Plot:LinearResponsePlot}
\end{figure}

In the late 2000s, \Abinit\ developed a utility---the U(J) Determination (\ujdet) protocol---designed to determine the Hubbard parameters based on Cococcioni and de Gironcoli's linear response method outlined in Section \ref{Linear Response}. In \Abinit\ versions 5 to 9.9, when activated, the protocol would serially introduce two perturbations---one of strength \abinput{pawujv} and the other of strength -1.0 $\times$ \abinput{pawujv}---to the subspace-uniform potential and harvest the resulting occupancy responses at the beginning and end of the two ensuing self-consistent cycles. The utility then had two points with which it could calculate the screened and unscreened response matrices, $\chi$ and $\chi_0$, defined as derivatives of occupation with respect to perturbation strength.

It is important to note that the positive value of \abinput{pawujv} was applied first, followed by its negative image. One would expect, then, that in performing two separate runs with the positive and negative values of \abinput{pawujv} should yield the same linear response.

This was not the case in \Abinit\ versions prior to 9.6.2. To demonstrate the error, we produced figure \ref{Plot:LinearResponsePlot} by performing $\beta$ perturbations on a particular system (ferromagnetic NiO) and monitoring the response from both the first and second \ujdet\ calculations, respectively. That is, we categorized perturbation-occupancy pairs according to their place in the queue in this double perturbation cycle. The screened response $\chi$ relaxes to a reasonably similar value regardless of its status as first or second calculation. The unscreened responses, however, are different depending on which perturbation is applied first. This discrepancy contributes to Hund's J parameters differing, in this case, by several eV. The same phenomenon was observed for the $\alpha$ perturbations contributing to the Hubbard U. The fact that the unscreened occupations differed depending on their place in the queue indicated that the perturbations were not being applied to the same initial ground state. Internal variables were not undergoing proper initialization, and so the second perturbation was inheriting information from the converged state of the preceding perturbative cycle.

\begin{figure*}[t]
    \centering
    \includegraphics[trim={0cm 0cm 9cm 0cm},clip,width=\linewidth]{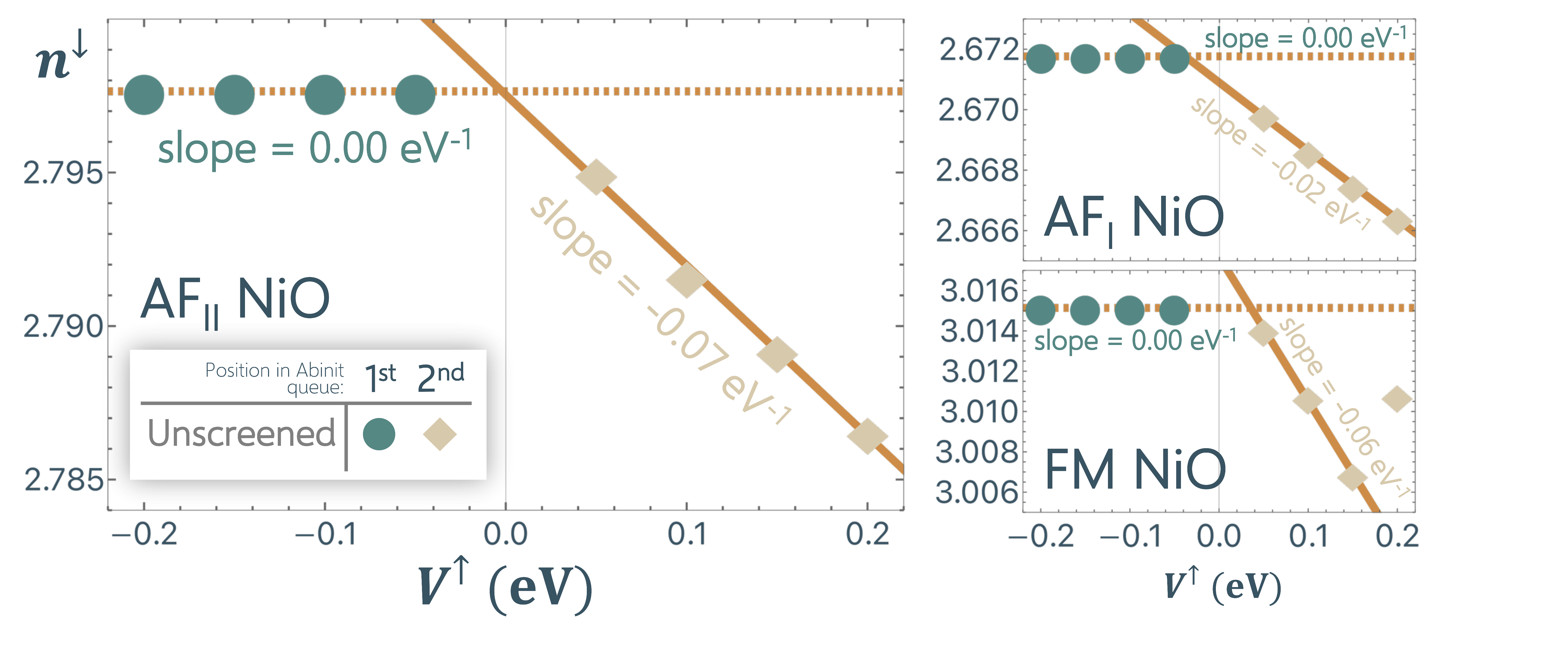}
    \caption{Comparison of unscreened spin-down occupation changes in response to a perturbation $\alpha$ applied exclusively to the spin-up $3d$ channel of a Ni atom on NiO of various magnetic orderings, where (\abicode{vsh}, \abicode{occ}) pairs are categorized according to their queue (first perturbation applied or second) in the full DFT cycle. We expect the slopes of these lines to be zero, but only the first \Abinit\ calculation fulfills this requirement.}
    \label{macro_uj2}
\end{figure*}

There are a few methods available to test which $\chi_0$ is the correct one. We know that, for the same system, where $\alpha=\beta$,
      
\begin{equation}\label{chi_0_alphabeta}
    \frac{dN_0}{d\alpha}=\frac{dM_0}{d\beta}.
\end{equation}
      
\noindent as was discussed in Section \ref{Linear Response}. This requisite is fulfilled only for the first applied perturbation (i.e., only the value supplied in \abinput{pawujv}, not its negative counterpart). Furthermore, we know that when we apply a potential perturbation to the spin-up channel only, the unscreened occupancy on the spin-down channel should not change. Applying a perturbation exclusively to the spin-up channel can be achieved by setting \abinput{macro\_uj}\texttt{=2} (the \abinput{macro\_uj} input parameter will be explained in Section \ref{Step2LRUJ}). We ran perturbations under this setting, applying perturbations to the spin-up channel only of a Ni atom and monitoring the change in unscreened occupancy on the down spin channel of the same Ni atom. The results of this inquiry, displayed in figure \ref{macro_uj2}, show that the unscreened occupancy on the spin-down channel remains constant only for the first applied perturbation, thereby corroborating the earlier conclusion that the second applied perturbation in the \Abinit\ cycle is unreliable. The silver lining for users of \ujdet\ prior to \Abinit\ 9.10.1 is that the first perturbation-occupancy data point is still salvageable.

\addcontentsline{toc}{section}{\spc\spc\spc\spc D: The Hubbard U parameter determination via \emph{\ujdet}}
\subsection*{\label{ujdet_calculation} Appendix D: The Hubbard U parameter determination via \emph{\ujdet}}

All subroutines constructing \Abinit's \ujdet\ utility—an abbreviation of ``Hubbard U and J determination"—are housed inside module \texttt{65\_paw/m\_paw\_uj.F90}. When the SCF cycle is complete, the same subroutine that launched the SCF cycle and allocated default variables for the \abicode{dtpawuj} type, \abicode{pawuj\_drive}, calls the subroutine \abicode{pawuj\_det}.

Once called, this subroutine calculates and prints the scalar Hubbard parameter for exclusively the perturbed atom using the two data points it has (i.e., the unperturbed occupancies and those of the one perturbation applied during its run). It is here that the program creates the NetCDF file with suffix \texttt{LRUJ.nc} for this perturbative run, filling it with all information that the \lruj\ post-processor will need to determine the choice Hubbard parameter in tandem with other perturbations. Before \Abinit\ wraps up its DFT run, however, the \ujdet\ algorithm switches to the matrix prescription for calculating the Hubbard parameters. In doing so, it proceeds to calculate all elements of the response matrices using the aforementioned (\abicode{vsh}, \abicode{luocc}) pairs in the following manner:

\begin{align}
    {\chi_{0_{t_it_j}}}&=\frac{{\abicode{luocc3}}_{t_i}-{\abicode{luocc1}}_{t_i}}{\abinput{diemix}\ ({\abicode{vsh3}}_{t_j}-{\abicode{vsh1}}_{t_j})} \label{chi0_abi} \\
    \chi_{t_it_j}&=\frac{{\abicode{luocc4}}_{t_i}-{\abicode{luocc2}}_{t_i}}{{\abicode{vsh4}}_{t_j}-{\abicode{vsh2}}_{t_j}} \label{chi_abi}
\end{align}

\noindent where $t_i,t_j$ are all atoms of the same species as the perturbed atom. 

\edit{These matrices will typically feature only one element: the diagonal element for which the perturbed atom $t_j$ is equal to the responding atom $t_i$. To extend this matrix and avail of the URES supercell facilities described below, one must tell \Abinit\ to monitor the occupancy responses on all atoms $t_i$ in the cell (see note on \abinput{symrel} in Section \ref{Step1LRUJ} for information on how to do this).} 

\edit{If going with the \abinput{symrel} procedure, it is furthermore recommended to set the input variable \abinput{pawprtvol}\texttt{=3} in order to have the occupancies of all Hubbard atoms printed out in the \texttt{.log} file. There, the (\abicode{vsh}, \abicode{luocc}) pairs will be listed for every atom $t_i$ after every SCF iteration. (The (\abicode{vsh4}, \abicode{luocc4}) pair is thus listed several times as the calculation converges; the relevant listing is, expectantly, the last one).}

\begin{figure*}[t]
    \begin{minipage}{\textwidth}
        \centering
        \includegraphics[trim={0.75cm 1cm 4cm 1.5cm},clip,width=1.0\textwidth]{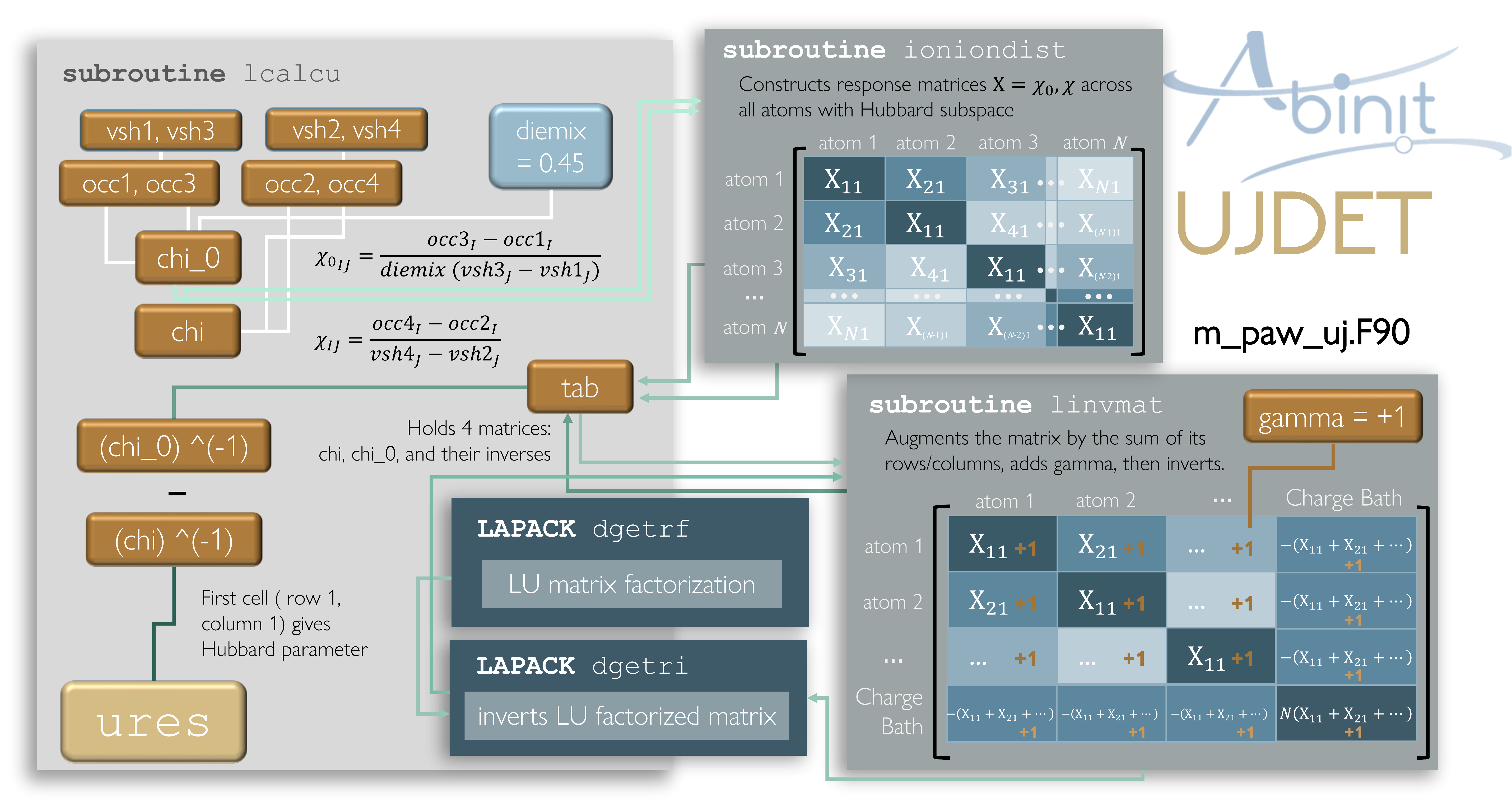}
        \caption{\edit{Flowchart illustrating manipulation of (\abicode{vsh}, \abicode{occ}) pairs to determine the final Hubbard parameter, \abicode{ures}, for a transition metal (TM) Hubbard subspace.}}
       \label{ujdet_utility_flowchart}
    \end{minipage}
\end{figure*}

\edit{Even in this case, since perturbations are always be applied to atom $t_j=1$, the only elements calculated via equations \ref{chi0_abi} and \ref{chi_abi} are those pertaining to $(t_i,t_j)=\{\ (1,1),(2,1),(3,1),...\ \}$. So when these matrices are subsequently funneled, via the mother Hubbard U subroutine \abicode{lcalcu}, to subroutine \abicode{ioniondist}, the lower triangular elements are completed via symmetry (i.e., $\textrm{X}_{JI}=X_{IJ}$, where $\textrm{X}=\chi_0,\chi$). The diagonal elements of the response matrices will all be set equal to $\textrm{X}_{11}$, as illustrated in figure \ref{ujdet_utility_flowchart}. Once completed, the matrices are returned to subroutine \abicode{lcalcu} and saved into variable \abicode{tab}, which holds four matrices: $\chi_0$, $\chi$, and their matrix inverses. It follows that \abicode{tab} is shuffled over to subroutine \abicode{linvmat}, which calculates the inverses of not the response matrices themselves, but treated matrices designed to speed up the convergence of the Hubbard parameters with respect to supercell size.}

These treatments are mentioned in the ``Further Considerations" section of reference \cite{cococcioni_linear_2005}, where it is posited that the perturbation on the Hubbard subspace would benefit from enhanced locality if charge neutrality in the response matrices was enforced, thereby isolating the perturbed atom from its periodic images, as one hopes to do using supercells. Following this understanding, \Abinit\ augments the response matrices with the negative of the sum of each row and each column, as illustrated in figure \ref{ujdet_utility_flowchart}.

This augmented matrix is, by definition, singular and thus non-invertible. To render the matrix invertible, an all-ones matrix is added to it, breaking its singularity. Note that this matrix is no longer equivalent to the input response matrices. However, as demonstrated in Appendix A.4 of reference \cite{linscott_thesis_2019}, the difference of the inverses of two non-invertible matrices—which is not possible mathematically—may be calculated indirectly by adding the same non-zero constant to each matrix element. This renders these matrices invertible, and the added constant is canceled when taking the difference of the two matrices.

Once prepped, the response matrices are funneled into the LAPACK routines \abicode{dgetrf} and \abicode{dgetri}, which respectively, LU factorize the matrices then invert them. These inverted matrices are then saved into the last two positions of \abicode{tab} and returned back to subroutine \abicode{lcalcu}. At last, the inverted response matrices are subtracted, then scaled by a factor called \abicode{signum} (\texttt{=1.0} for the Hubbard U and \texttt{=-1.0} for the Hund's J). The first element of that object then (row 1, column 1), in eV, is found to be the long-awaited Hubbard parameter.

The \ujdet\ utility does not stop here, though. The ``Further Considerations" section of reference \cite{cococcioni_linear_2005} considers a hypothetical extrapolation scheme speculatively designed to converge much more quickly the Hubbard parameter with respect to supercell size. The number of Hubbard subspaces in a supercell corresponds linearly with the response matrix dimension. But intuition suggests that the occupancy effect of the perturbed subspace will attenuate with distance; that is, the matrix elements of the nearest neighbor atoms to that perturbed will feature most prominently in the determination of the Hubbard parameter, and those least neighborly to the perturbed atom will undergo small, even negligible, changes in occupancy, rendering their influence negligible. \Abinit\ developers took these further considerations to heart by incorporating an extrapolation scheme, wherein the response matrix elements of the primitive unit cell are used to fill out the response matrix elements of a supercell. Concisely, in \Abinit's \ujdet\ utility, the off-diagonal elements of the primitive cell response matrices are multiplied by the number of next-nearest neighbor (NNN) Hubbard atoms in the primitive cell and divided by the number of Hubbard atoms in NNN shell in the supercell. These supercell response matrices are then inverted following the same procedure as above to approximate the Hubbard parameters for subspaces in supercells of increasing size.

\addcontentsline{toc}{section}{\spc\spc\spc\spc E: What AbiPy can do with a \lruj\ output file}
\subsection*{\label{AbiPy} Appendix E: What AbiPy can do with a \lruj\ output file}

This mixing constant-corrected linear response plot can be easily generated through the \href{https://abinit.github.io/abipy/index.html}{AbiPy} python package; its version 0.9.7 is able to read in results from one's chosen \texttt{lruj.out} output file and visualize its results. To avail of this functionality, educe a python script in the same directory as that containing your \texttt{lruj.out} output. Begin the python script by importing the \texttt{LrujResults} function from the AbiPy package:

\vspace{0.5cm}
\textcolor{comment}{\texttt{\#\!/usr/bin/env python}}

\texttt{\textcolor{pyimport}{from} abipy.electrons.lruj \textcolor{pyimport}{import} LrujResults}
\vspace{0.5cm}

\noindent Import the \texttt{lruj.out} file using the following line.

\vspace{0.5cm}
\texttt{lr = LrujResults.from\_file(\textcolor{pyfile}{"\textit{path\_to\_file}/lruj.out"})}
\vspace{0.5cm}

\noindent The plot function may then be summoned with

\vspace{0.5cm}
\texttt{lr.plot\textcolor{pyoptions}{(\textit{ax}, \textit{degrees}, \textit{inset}, \textit{insetdegree}, \textit{insetlocale}, \textit{ptcolor0},}}

\texttt{\spc\spc\textcolor{pyoptions}{\textit{ptcolor}, \textit{gradcolor1}, \textit{gradcolor2}, \textit{ptitle}, \textit{fontsize})}}

\vspace{0.5cm}

\noindent where the arguments listed in blue are keywords to tailor particular characteristics of the ensuing plot. These options are described in more detail in table \ref{Tab:AbiPy_plot_options}.

Other AbiPy tools to better accommodate the linear response process in \Abinit\ are currently in the works. For example, under development is a suite of functions that aim to facilitate visualization of the convergence of the Hubbard parameters with respect to supercell size. Keep an eye on forthcoming releases of \href{https://abinit.github.io/abipy/index.html}{AbiPy} for such developments.

\setlength\tabcolsep{5pt}
\setlength{\arrayrulewidth}{0.4mm}
\def\arraystretch{1.5}
\begin{longtable}{P{0.15\linewidth}|P{0.15\linewidth}|P{0.3\linewidth} | >{\setlength{\baselineskip}{0.65\baselineskip}} L{0.3\linewidth} |}

\rowcolor[HTML]{668A82}[\overhang] 
\multicolumn{4}{c}{\textcolor{white}{\texttt{LrujResults plot} utility optional arguments}} \\ \hline

\rowcolor[HTML]{395464}[\overhang] 
\textcolor{white}{Argument} & \textcolor{white}{Default} & \textcolor{white}{Other Options} & \textcolor{white}{Description} \\ \hline

\rowcolor[HTML]{EEF2F4}[\overhang]
{\textcolor{pyoptions}{\texttt{ax}}} & \texttt{\textcolor{pyfunction}{None}} &  \texttt{\textcolor{pyfunction}{ax}}  &  {\footnotesize Optional axes argument. If \texttt{\textcolor{pyfunction}{None}}, a new plot is generated. If \texttt{\textcolor{pyfunction}{ax}}, the figure without the axes is returned. Useful for generation of a grid of plots.}  \\ \hline

\rowcolor[HTML]{EEF2F4}[\overhang]
{\textcolor{pyoptions}{\texttt{degrees}}} & \texttt{\textcolor{pyfile}{\texttt{"all"}}} &  List of integers $i$ such that $0<i<$ \texttt{maximum degree} & {\footnotesize Degrees of polynomial regressions to be included in the plot, provided as a Python list of integers (e.g., \texttt{$[1,2,3]$}). The maximum integer allowed is \texttt{maximum degree}, which is read in via the \texttt{lruj.out} output file }   \\ \hline

\rowcolor[HTML]{EEF2F4}[\overhang]
{\textcolor{pyoptions}{\texttt{inset}}} & \texttt{\textcolor{pyfunction}{True}} & \texttt{\textcolor{pyfunction}{False}} & {\footnotesize Option to print inset with response information (i.e., values of $\chi_0$, $\chi$, the Hubbard parameter and their respective errors in units of eV). If \texttt{\textcolor{pyfunction}{True}}, information is printed for the linear regression case in the lower left corner of the plot (by default; see \textcolor{pyoptions}{\texttt{insetdegree}} and \textcolor{pyoptions}{\texttt{insetlocale}} options to tailor). If \texttt{\textcolor{pyfunction}{False}}, no inset is included.} \\ \hline

\rowcolor[HTML]{EEF2F4}[\overhang]
{\textcolor{pyoptions}{\texttt{insetdegree}}} & \texttt{1} &  Any integer $\in$ {\textcolor{pyoptions}{\texttt{degrees}}}  & {\footnotesize Polynomial degree of printed response information appearing in inset.}  \\ \hline

\rowcolor[HTML]{EEF2F4}[\overhang]
{\textcolor{pyoptions}{\texttt{insetlocale}}} & \texttt{\textcolor{pyfile}{"lower left"}} &  \texttt{\textcolor{pyfile}{"upper right"}},  \texttt{\textcolor{pyfile}{"center left"}}, \texttt{\textcolor{pyfile}{"lower center"}}, \texttt{\textcolor{pyfile}{"center"}}, corresponding integers 0-10, etc.  & {\footnotesize Position of inset containing response information in standard format for \texttt{matplotlib} legend locations. See \href{https://matplotlib.org/2.0.2/api/colors_api.html}{\texttt{matplotlib} documentation} for all options.} \\ \hline

\rowcolor[HTML]{EEF2F4}[\overhang]
{\textcolor{pyoptions}{\texttt{ptcolor0}}} & \texttt{\textcolor{pyfile}{"k"}} & \texttt{\textcolor{pyfile}{"r"}}, \texttt{\textcolor{pyfile}{"blue"}}, \texttt{\textcolor{pyfile}{"FF6E42"}}, \texttt{(0.1,0.9,0.54)}, \texttt{\textcolor{pyfile}{"0.75"}}, etc. & {\footnotesize Color of unscreened response data point markers in any standard \texttt{matplotlib} color format (see \href{https://matplotlib.org/2.0.2/api/colors_api.html}{\texttt{matplotlib} documentation} for all formatting options.  Default color is black. Markers themselves are open circles \faCircleO\ (immutably so, for now).}   \\ \hline

\rowcolor[HTML]{EEF2F4}[\overhang]
{\textcolor{pyoptions}{\texttt{ptcolor}}} & \texttt{\textcolor{pyfile}{"k"}} & \texttt{\textcolor{pyfile}{"r"}}, \texttt{\textcolor{pyfile}{"blue"}}, \texttt{\textcolor{pyfile}{"FF6E42"}}, \texttt{(0.1,0.9,0.54)}, \texttt{\textcolor{pyfile}{"0.75"}}, etc. & {\footnotesize Color of screened response data point markers in any standard \texttt{matplotlib} color format (see \href{https://matplotlib.org/stable/api/_as_gen/matplotlib.pyplot.scatter.html}{\texttt{matplotlib} documentation} for all formatting options. Default color is black. Markers themselves are filled circles \faCircle\ (immutably so, for now).} \\ \hline

\rowcolor[HTML]{EEF2F4}[\overhang]
{\textcolor{pyoptions}{\texttt{gradcolor1}}} & \texttt{\textcolor{pyfile}{"\#3575D5"}} & Hexadecimal (HEX) code for any color & {\footnotesize Line color of the lowest polynomial degree to be plotted (i.e., the smallest integer input via \textcolor{pyoptions}{\texttt{degrees}} argument). This color, in addition to that of \textcolor{pyoptions}{\texttt{gradcolor2}} will inform the line colors of the intermediate polynomial degrees in a linear gradient fashion. Must be entered as a HEX code (six characters preceded by a `\#'). The default color is \textcolor{NewBlue}{dark blue}.} \\ \hline

\rowcolor[HTML]{EEF2F4}[\overhang]
{\textcolor{pyoptions}{\texttt{gradcolor2}}} & \texttt{\textcolor{pyfile}{"\#FDAE7B"}} & Hexadecimal (HEX) code for any color & {\footnotesize Line color of the highest polynomial degree to be plotted (i.e., the largest integer input via \textcolor{pyoptions}{\texttt{degrees}} argument). Must be entered as a Hex code (six characters preceded by a `\#'). The default color is \textcolor{salmon}{salmon pink}.} \\ \hline

\rowcolor[HTML]{EEF2F4}[\overhang]
{\textcolor{pyoptions}{\texttt{ptitle}}} & \texttt{\textcolor{pyfile}{"Linear Response for atom <\textit{pawujat}>"}} &  Any string  &  {\footnotesize Title of plot. Incorporates input value of \texttt{pawujat} by default. For no title, put \texttt{\textcolor{pyfile}{""}}.}  \\ \hline

\rowcolor[HTML]{EEF2F4}[\overhang]
{\textcolor{pyoptions}{\texttt{fontsize}}} & \texttt{12} & Any integer $>0$ & {\footnotesize Font size in point (pt) units of the plot legend. }   \\ \hline

\caption{\label{Tab:AbiPy_plot_options}Optional arguments and their possible values available in the \texttt{plot} function implemented in the \texttt{LrujResults} function of the AbiPy package.}
\end{longtable}

\newpage
\nocite{*}

\bibliography{Manuscript}

\begin{thebibliography}{10}
\bibitem{footnote1}
\Abinit\ 9.6.2 did not originally have functionality implemented to apply a $\beta$-perturbation as it is described in Section \ref{Linear Response} to acquire the Hund's J. The version of \Abinit\ used for these calculations was modified locally to do so. These local modifications later served as the blueprint for the official Hund's J implementation in \Abinit\ Version 9.10.1.
\end{thebibliography}

\end{document}